\documentclass[12pt]{iopart}
\usepackage{iopams} 

\usepackage{graphicx}
\usepackage{color}
\usepackage{braket}
\usepackage{xspace}
\usepackage[pagewise]{lineno}

\usepackage{bm}
\usepackage{tabularray}
\usepackage{xspace}
\usepackage[hidelinks]{hyperref}
\newcommand{\EF}{$E_\mathrm{F}$\xspace}
\newcommand{\Uf}{$\mathrm{U}~5f$\xspace}
\newcommand{\Uff}{$\mathrm{U}~4f$\xspace}
\newcommand{\orb}[2]{$\mathrm{ #1 } ~ #2 $\xspace}
\newcommand{\hn}[1]{$h\nu#1~\mathrm{eV}$\xspace}
\newcommand{\hnk}[1]{$h\nu#1~\mathrm{keV}$\xspace}
\newcommand{\EB}[1]{$E_{\mathrm{B}} #1~\mathrm{eV}$\xspace}

\newcommand{\pnt}[1]{$\mathrm{#1}$\xspace}
\newcommand{\Gm}{$\mathrm{\Gamma}$\xspace}

\newcommand{\Udf}{$\mathrm{U}~4d-5f$\xspace}

\newcommand{\Uedge}[1]{\texorpdfstring{$\mathrm{U}~{#1}$}{U #1}}
\newcommand{\nf}[1]{$n_{5f}{#1}$\xspace}
\newcommand{\Unf}[1]{$\mathrm{U}~5f^{#1}$\xspace}

\newcommand{\UTeTe}{\texorpdfstring{$\mathrm{UTe_2}$}{UTe2}\xspace}
\newcommand{\ThTeTe}{$\mathrm{ThTe_2}$\xspace}
\newcommand{\UCoGe}{$\mathrm{UCoGe}$\xspace}
\newcommand{\URhGe}{$\mathrm{URhGe}$\xspace}
\newcommand{\UGe}{$\mathrm{UGe_{2}}$\xspace}
\newcommand{\URuSi}{$\mathrm{URu_{2}Si_{2}}$\xspace}
\newcommand{\UPdAl}{$\mathrm{UPd_{2}Al_{3}}$\xspace}
\newcommand{\UPt}{$\mathrm{UPt_{3}}$\xspace}
\newcommand{\UB}{$\mathrm{UB_{2}}$\xspace}
\newcommand{\UAl}{$\mathrm{UAl_{3}}$\xspace}
\newcommand{\UGa}{$\mathrm{UGa_{2}}$\xspace}
\newcommand{\UPd}{$\mathrm{UPd_{3}}$\xspace}

\newcommand{\GGAU}{GGA+$U$\xspace}

\newcommand{\ea}{\textit{et al}\xspace}

\begin{document}

\review[Probing the electronic structure of $\mathrm{UTe}_2$ with ARPES and high-energy spectroscopy]{Probing the electronic structure of $\mathrm{UTe}_2$ with ARPES and high-energy spectroscopy}

\author{Shin-ichi Fujimori}

\address{Materials Sciences Research Center, Japan Atomic Energy Agency, Sayo, Hyogo 679-5148, Japan}
\ead{fujimori@spring8.or.jp}

\begin{abstract}
$\mathrm{UTe}_2$ has emerged as one of the most intensively studied strongly correlated materials in recent years owing to its unconventional superconductivity and the possible realization of a spin-triplet, topologically nontrivial pairing state.
A central open issue concerns the nature of the $\mathrm{U}\,5f$ electrons and their participation in low-energy quasiparticle states.
In this review, we summarize recent spectroscopic studies of $\mathrm{UTe}_2$ using momentum-resolved angle-resolved photoemission spectroscopy (ARPES) and element- and configuration-sensitive X-ray probes, including X-ray absorption spectroscopy (XAS), X-ray absorption near-edge structure (XANES), X-ray magnetic circular dichroism (XMCD), resonant X-ray emission spectroscopy (RXES), and resonant inelastic X-ray scattering (RIXS).
A key finding is the pronounced technique dependence of the inferred electronic structure.
We synthesize the present spectroscopic picture by integrating these results with modern electronic-structure calculations such as density functional theory plus dynamical mean-field theory (DFT+DMFT).
These findings support an intermediate-valence ground state with significant admixture of $5f^2$ and $5f^3$ configurations in $\mathrm{UTe}_2$, and they delineate key experimental and theoretical benchmarks needed to connect the normal-state electronic structure to the superconducting mechanism.
\end{abstract}

\tableofcontents

\section{\label{sec:introduction}Introduction}
The discovery of superconductivity in uranium ditelluride (\UTeTe) has triggered intense and wide-ranging research efforts in recent years~\cite{UTe2_sc,UTe2_Aoki}.  
Beyond its unconventional superconducting properties, \UTeTe has emerged as a prototypical strongly correlated system where the nature of the \Uf electrons plays a central role.  
As a result, a broad array of spectroscopic studies has been carried out to elucidate its electronic structure across multiple energy scales.
Angle-resolved photoemission spectroscopy (ARPES) has been extensively employed to investigate the momentum-resolved band structure and Fermi surface, thereby providing direct access to low-energy quasiparticle excitations.  
In \UTeTe, ARPES has revealed several important features of the low-energy band structure~\cite{UTe2_ARPES,UTe2_ARPES_Wray,UTe2_ARPES_Wray2}; however, its relationship to the Fermi-surface topology inferred from de Haas--van Alphen (dHvA) oscillation experiments~\cite{UTe2_dHvA} and from modern theoretical calculations, such as density functional theory (DFT) and DFT plus dynamical mean-field theory (DFT+DMFT), remains unresolved.

In parallel, a variety of X-ray spectroscopic techniques have been applied to probe local electronic states, valence configurations, and orbital degrees of freedom of the \Uf electrons.  
While these approaches provide valuable and complementary information, their interpretations have not always converged on a unified picture of the $5f$ electronic structure of \UTeTe.  
Reconciling these different perspectives remains an open and important challenge.

These unresolved spectroscopic issues are particularly important because the same low-energy electronic states are expected to constrain the superconducting pairing state of \UTeTe.
The Fermi-surface topology, magnetic fluctuations, and local electronic structure are closely related to the superconducting order parameter of \UTeTe, which remains an active and not yet fully settled issue.
Early observations of the extremely large and anisotropic upper critical field, reentrant superconductivity, and signatures of chiral or odd-parity superconductivity have motivated interpretations in terms of spin-triplet and possibly topological pairing~\cite{UTe2_sc,UTe2_Aoki_review,Jiao_UTe2_Nature2020,UTe2_OddParity_Gu,UTe2_QPI_Wang}.
Inelastic-neutron-scattering experiments revealed pronounced low-energy incommensurate magnetic fluctuations rather than simple ferromagnetic fluctuations, indicating that the magnetic excitation spectrum is strongly constrained by the underlying electronic structure and possible nesting vectors of the Fermi surface~\cite{DMFT_Duan}.
Recent DFT+DMFT calculations further suggest that the same correlation effects that generate heavy quasiparticle bands and modify the Fermi-surface topology also influence the momentum structure of the magnetic excitations~\cite{DFTDMFT_Halloran}.
At the same time, ultrasound, Knight-shift, penetration-depth, specific-heat, thermodynamic, and theoretical studies continue to discuss the order-parameter symmetry, gap nodes, time-reversal-symmetry breaking, and the roles of spin--orbit coupling, magnetic fluctuations, sample quality, and field-induced phases~\cite{UTe2_SingleComponent_Theuss,UTe2_Knight_Fujibayashi,UTe2_Chiral_Ishihara_NatCommun2023,UTe2_SC_sym,UTe2_PRB112_054510}.
In this context, the recent study of next-generation ultraclean crystals by Wu \ea has emphasized that sample purity can strongly enhance the superconducting response and clarify intrinsic high-field behavior, underscoring that disorder and residual scattering must also be considered when relating electronic structure to superconductivity~\cite{UTe2_Ultraclean_PNAS}.
Thus, while \UTeTe is a strong candidate for odd-parity spin-triplet superconductivity, its precise superconducting symmetry should be considered together with the unresolved Fermi-surface topology and the momentum-dependent magnetic fluctuations.

The controversies over the \Uf electronic structure parallel the long-standing debates surrounding the hidden-order compound \URuSi~\cite{URu2Si2_review_2020,URu2Si2_Palstra}, where the nature of the \Uf electrons has remained controversial for decades.  
In that case, different experimental probes and theoretical approaches have led to conflicting pictures of the ground-state electronic structure, and establishing an electronic-structure model has proven challenging even at present.
More generally, in uranium-based materials the dichotomy between itinerant and localized descriptions of the $5f$ states has often served as a useful, albeit simplified, conceptual framework for organizing these disparate results.  
While this binary classification cannot fully capture the dual and energy-dependent character of $5f$ electrons, it continues to shape interpretations of low-energy phenomena and underlies many of the central controversies in the field.

In this context, \UTeTe appears to reside near a delicate boundary between localized and itinerant descriptions of the \Uf electrons.
While some experimental and theoretical studies emphasize signatures consistent with a nearly localized $5f^{2}$ configuration~ \cite{UTe2_ARPES_Wray,UTe2_O_RIXS,UTe2_M_RIXS}, others point to a mixed-valence state with an average $5f$ occupation significantly deviating from an integer value~ \cite{UTe2_L3_XANES,UTe2_core,UTe2_XMCD,UTe2_RXES,UTe2_U4fDMFT}.
Crucially, such mixed-valence behavior does not necessarily imply strong band-like itinerancy, but may instead reflect local valence fluctuations arising from the near degeneracy of $5f^{2}$ and $5f^{3}$ configurations.
Understanding how these distinct aspects are selectively highlighted by different experimental probes, and how they can be reconciled within a unified physical framework, is essential for interpreting the electronic structure of \UTeTe and its unconventional superconductivity.

In this review, we present an overview of ARPES and X-ray spectroscopic studies of \UTeTe, with an emphasis on how different techniques probe distinct aspects of the uranium $5f$ electronic structure.  
While a comprehensive review of its superconducting and thermodynamic properties has been provided by Aoki \ea~\cite{UTe2_Aoki_review}, we focus here on the normal-state electronic structure revealed by momentum-resolved ARPES measurements and on its connection to X-ray spectroscopies.  
By comparing and integrating results from these complementary approaches, we aim to clarify the valence character, degree of itinerancy, and hybridization of the $5f$ states in \UTeTe, and to place this material within the broader landscape of correlated uranium compounds.

This review is organized as follows.  
Section~\ref{sec:electronic_structure} summarizes the theoretical and experimental understanding of the electronic structure of \UTeTe.  
Section~\ref{sec:ARPES} reviews recent ARPES studies.  
Section~\ref{sec:Xray} discusses X-ray spectroscopic investigations.  
Section~\ref{sec:discussion} integrates these findings and provides concluding remarks.
Finally, the summary and outlook are presented in Section~\ref{sec:summary}.

\section{\label{sec:electronic_structure}Electronic Structure of \UTeTe}
\subsection{\label{sec:crystal_BZ}Crystal structure and Brillouin zone}
\begin{figure*}[tb]
	\centering
	\includegraphics[scale=0.50]{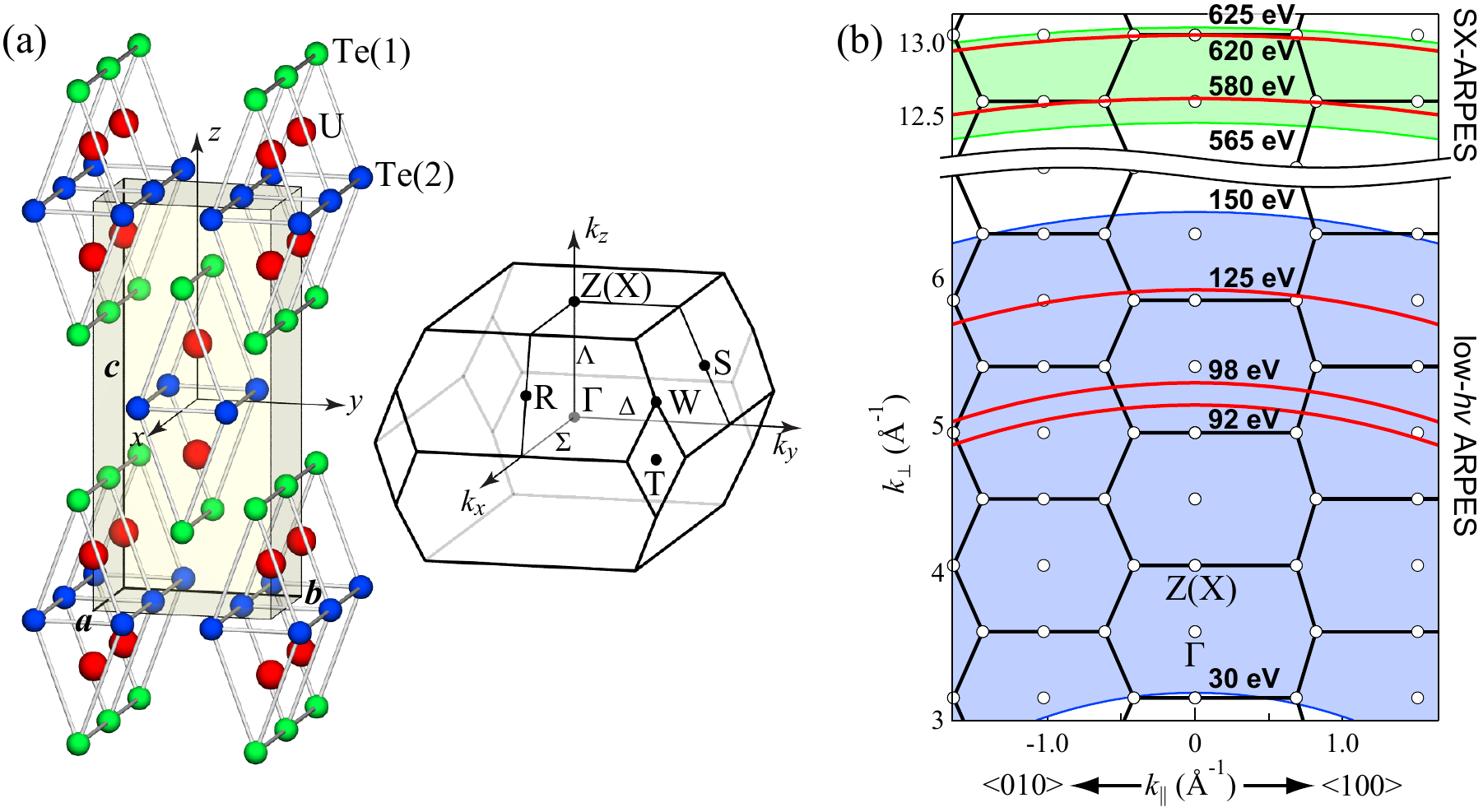}
	\caption{
	Crystal structure and Brillouin zone of \UTeTe.
	(a) Crystal structure (left) and Brillouin zone of \UTeTe.
	(b) ARPES cuts in multi-Brillouin-zone momentum space along the $\langle 100\rangle$ ($k_a$) and $\langle 010\rangle$ ($k_b$) directions.
	Panel (a) is reprinted from Fig.~1 of Fujimori \etal, 2019~\cite{UTe2_ARPES}.
	}
	\label{fig:UTe2_BZ}
\end{figure*}

The crystal structure and Brillouin zone (BZ) of \UTeTe are shown in Fig.~\ref{fig:UTe2_BZ}.
\UTeTe crystallizes in a body-centered orthorhombic structure with the symmorphic space group $Immm$ (No. 71, $D^{25}_{2h}$).
The conventional unit cell contains four formula units, and the tellurium atoms occupy two inequivalent crystallographic sites, Te(1) and Te(2).
Because the crystallographic $a$, $b$, and $c$ axes are not equivalent, the uranium sublattice forms anisotropic structural motifs that provide a natural structural basis for the direction-dependent electronic structure discussed below.
Note that the natural cleavage surface of \UTeTe is the (0--11) surface, or the symmetry-related $(011)$ surface.
On the other hand, \UTeTe can also be cleaved along the principal $(001)$ plane, and most ARPES studies have been performed on this plane.

For comparison between ARPES measurements and band-structure calculations, the BZ convention must be specified carefully.
In this review, momenta are expressed with respect to the conventional orthorhombic reciprocal-lattice vectors, and the high-symmetry-point notation follows the convention in Fig.~\ref{fig:UTe2_BZ}(a).
The point $(0,0,1)$ is denoted as either the \pnt{Z} or \pnt{X} point in the literature; here, it is referred to as the \pnt{Z} point.
The body-centered translation gives a BZ geometry different from that of a simple orthorhombic lattice, which should be taken into account when assigning ARPES features to calculated bulk bands.

Figure~\ref{fig:UTe2_BZ}(b) defines the multi-BZ ARPES cuts along the $\langle 100\rangle$ ($k_a$) and $\langle 010\rangle$ ($k_b$) directions, assuming that the principal $(001)$ plane is the cleaving plane.
For \UTeTe, both soft X-ray ARPES (SX-ARPES, \hn{=565-625})~\cite{UTe2_ARPES} and low-photon-energy ($h\nu$) ARPES (\hn{=30-150})~\cite{UTe2_ARPES_Wray} have been reported, and their Brillouin-zone coverage is also shown in the figure.
These cuts are referred to in Sec.~\ref{sec:ARPES}.

\subsection{\label{sec:TheoryFS}Theoretical Fermi surface}
\begin{table*}
\centering
\scalebox{0.8}{
\begin{tblr}{
    hline{1,Z} = {0.1em},
    hline{2} = {0.05em},
    colspec = {l l l c},
    cell{1}{1-Z} = {font=\bfseries, c},
}
Method		& Interaction parameters ($U$, $J$)	& Temperature		& Reference \\
GGA+$U$		& $U = 0$--$2.0$~eV, $J = 0$~eV		& --				& \cite{UTe2_GGAU}\\
DFT+DMFT	& $U = 8.0$~eV, $J = 0.6$~eV		& 10~K, 200~K		& \cite{DFTDMFT_Xu}\\
DFT+DMFT	& $U = 6.0$~eV, $J = 0.57$~eV		& 10~K, 100~K		& \cite{UTe2_ARPES_Wray}\\
DFT+DMFT	& $U = 6.0$~eV, $J = 0.57$~eV		& 10~K, 100~K		& \cite{DMFT_Duan}\\
DFT+$U$(ED)	& $U = 3.0, 6.0$~eV, $J = 0.51$~eV		& --				& \cite{UTe2_Shick2}\\
DFT+DMFT	& $U = 8.0$~eV, $J = 0.6$~eV		& 11~K				& \cite{DFTDMFT_Choi}\\
DFT+DMFT	& $U = 6.0$~eV, $J = 0.57$~eV		& 116~K, 232~K, 580~K	& \cite{DFTDMFT_Halloran}\\
LQSGW+DMFT	& $U = 3.29$~eV, $J = 0.24$~eV		& 25~K, 900~K		& \cite{UTe2_LQSGWDMFT2,UTe2_LQSGWDMFT1}\\
DFT+DMFT	& $U = 3.0$~eV, $J = 0.59$~eV		& 300~K				& \cite{UTe2_U4fDMFT}\\
\end{tblr}
}
\caption{Summary of electronic structure calculation studies on \UTeTe. 
The computational methods, Coulomb interaction ($U$) and Hund coupling ($J$) parameters, and calculation temperatures are listed for comparison.}
\label{calc}
\end{table*}
\begin{figure*}[tb]
	\centering
	\includegraphics[scale=0.45]{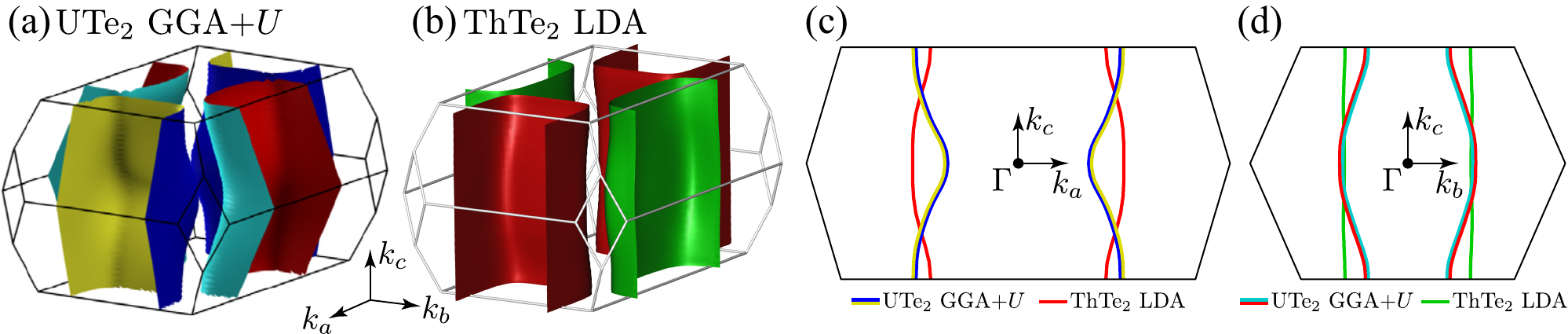}
	\caption{Calculated Fermi surfaces of \UTeTe.
	(a) Fermi surface calculated using GGA+$U$ with $U=2~\mathrm{eV}$.
	(b) Fermi surface calculated for \ThTeTe, corresponding to the localized-$5f^2$ limit.
	(c) Comparison in the $k_a$--$k_c$ plane.
	(d) Comparison in the $k_b$--$k_c$ plane.
	Panel (a) is reprinted from \cite{UTe2_GGAU} with permission, Copyright (2019) by the American Physical Society.
}
	\label{fig:ThTe2_UTe2}
\end{figure*}
Following the discovery of superconductivity in \UTeTe, a number of band-structure calculations based on density functional theory (DFT), including the local density approximation (LDA) and generalized gradient approximation (GGA), were performed~\cite{UTe2_Aoki, UTe2_ARPES}.
These early calculations predicted \UTeTe to be a narrow-gap semiconductor or semimetal with a very small carrier density, in apparent contradiction with its experimentally established metallic behavior.
To resolve this discrepancy, a variety of theoretical approaches beyond conventional DFT have subsequently been explored.

Table~\ref{calc} summarizes such theoretical studies reported for \UTeTe.  
Among them, DFT+$U$ calculations have been used to incorporate the strong on-site Coulomb interaction among the \Uf electrons~\cite{UTe2_GGAU,UTe2_Shick2}.
Because static DFT+$U$ approaches cannot capture the formation of heavy quasiparticle bands near the Fermi level, several groups have also performed calculations based on dynamical mean-field theory (DMFT)~\cite{DFTDMFT_Xu, UTe2_ARPES_Wray, DMFT_Duan, DFTDMFT_Choi, DFTDMFT_Halloran, UTe2_LQSGWDMFT2, UTe2_LQSGWDMFT1, UTe2_U4fDMFT}.  
Within these frameworks, the effective Coulomb interaction $U$ and Hund's coupling $J$ play decisive roles in determining the electronic structure and resulting Fermi surface.  

Depending on the value of $U$, these calculations predict qualitatively different electronic structures.
For moderate interaction strengths, $U = 2$--$3~\mathrm{eV}$, the \Uf states retain a partially itinerant character and contribute directly to the formation of the Fermi surface~\cite{UTe2_GGAU, UTe2_LQSGWDMFT2, UTe2_LQSGWDMFT1, UTe2_U4fDMFT, UTe2_Shick2, UTe2_Shick1}.  
Ishizuka \ea performed GGA+$U$ calculations and demonstrated that the Fermi-surface topology depends sensitively on the choice of the Coulomb parameter $U$~\cite{UTe2_GGAU}.  
Figure~\ref{fig:ThTe2_UTe2}(a) shows the Fermi surface obtained for $U=2~\mathrm{eV}$.
This topology agrees well with that inferred from dHvA oscillation experiments, as discussed in Sec.~\ref{sec:ExpFS}.

By contrast, for relatively strong interactions, $U = 6$--$8~\mathrm{eV}$, the \Uf states are separated into occupied lower-Hubbard-like and unoccupied upper-Hubbard-like bands, while the conduction bands and Fermi surface are primarily derived from \orb{U}{6d} and \orb{Te(2)}{5p} orbitals~\cite{UTe2_ARPES_Wray, DFTDMFT_Xu, DFTDMFT_Choi}.  
In this regime, the calculated band structure and Fermi surface closely resemble those of hypothetical \ThTeTe constructed in the same crystal structure as \UTeTe, which serves as a localized-$5f^2$ model for \UTeTe~\cite{UTe2_Harima}, even though actual \ThTeTe does not adopt this structure~\cite{DEye_ThTe_1954}.
This hypothetical Fermi surface is characterized by two orthogonal quasi-one-dimensional sheets extending along the $k_c$ direction, as shown in Fig.~\ref{fig:ThTe2_UTe2}(b).  

It should be noted that, in this strong-coupling regime, the overall Fermi-surface topology resembles that obtained from GGA+$U$ calculations with $U=2~\mathrm{eV}$.
The detailed shape along the $k_c$ direction, however, differs substantially.
As shown in Figs.~\ref{fig:ThTe2_UTe2}(c) and (d), although both calculations yield two orthogonal quasi-one-dimensional Fermi surfaces, their $k_c$ dependences differ markedly.
In particular, in the $k_a$--$k_c$ plane, the GGA+$U$ result exhibits a pronounced ``neck'' structure near the \Gm point, whereas the LDA calculation for \ThTeTe\ yields a more expanded and weakly modulated contour.
In addition, their band structures differ substantially, as shown in Fig.~\ref{fig:UTe2_ARPES_EF} and discussed further in Sec.~\ref{sec:nearEF}.
These distinctions are essential for disentangling the contribution of the $\mathrm{U}~5f$ states to the Fermi surface, as discussed further in Sec.~\ref{sec:ExpFS}.  

Regarding the strength of $U$, within DFT+$U$ or DFT+DMFT-based approaches, $U = 2$--$5~\mathrm{eV}$ is commonly used for uranium metals and oxides~\cite{UN_LDAU,LDADMFT_U}, whereas $U \gtrsim 6~\mathrm{eV}$ is likely too large for metallic \UTeTe, since photoemission and inverse-photoemission measurements suggest $U \simeq 4.6~\mathrm{eV}$ even for localized $\mathrm{UO}_2$~\cite{UO2_XPS_BIS}.

\subsection{\label{sec:ExpFS}Experimental Fermi surface}
\begin{figure*}
	\centering
	\includegraphics[scale=0.5]{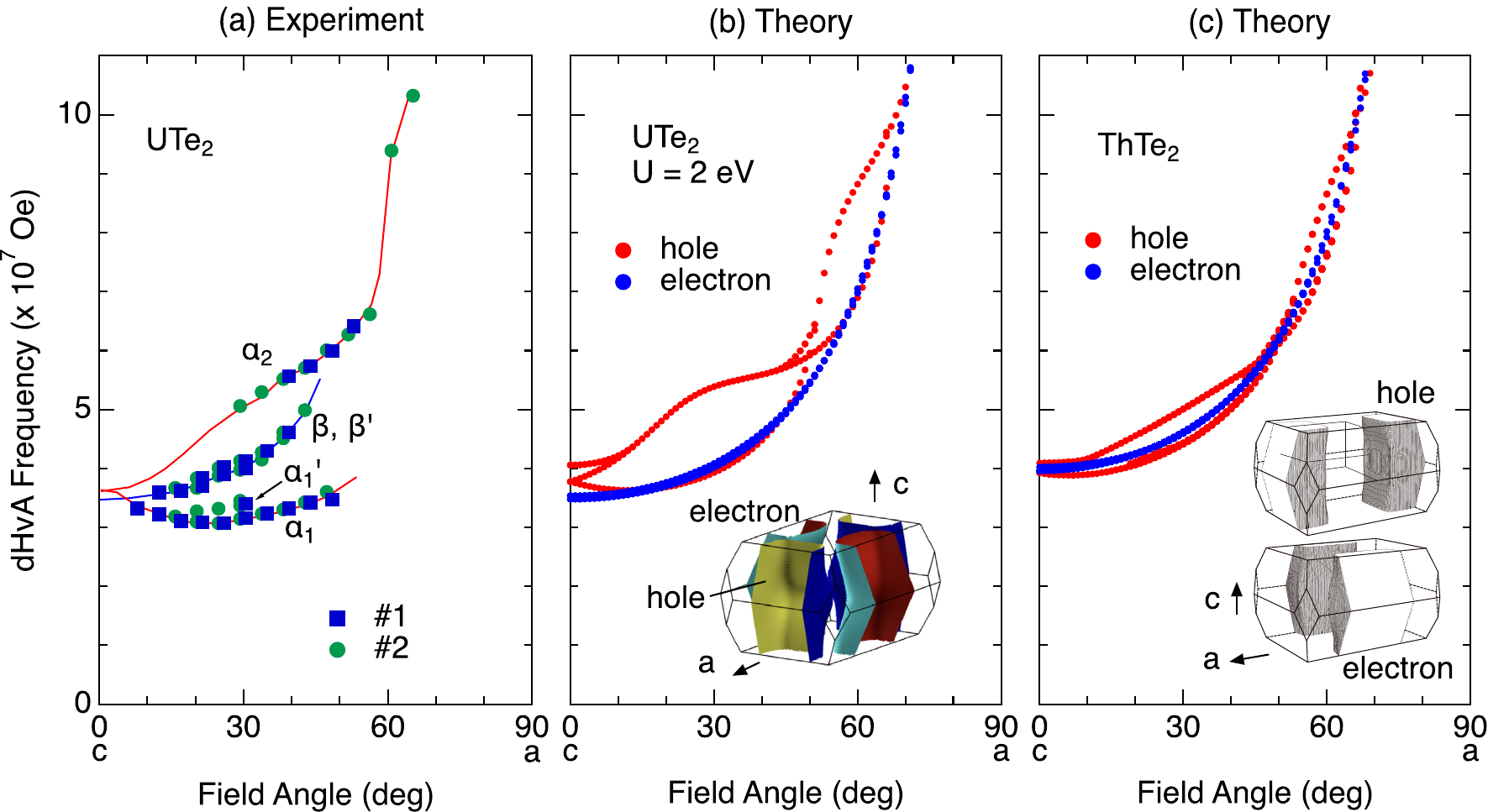}
	\caption{The result of dHvA experiment of \UTeTe.
	(a) Angular dependence of the dHvA frequency in \UTeTe.
	(b), (c) Angular dependence of the dHvA frequency calculated by \GGAU ($U=2~\mathrm{eV}$) in \UTeTe and LDA for \ThTeTe.
	Reproduced from Fig. 5 of Aoki \etal, 2022~\cite{UTe2_dHvA} under \href{https://creativecommons.org/licenses/by/4.0/}{CC BY 4.0}.
}
	\label{fig:UTe2_dHvA}
\end{figure*}
The Fermi-surface topology of \UTeTe\ has been investigated using several experimental techniques.
Because quantum oscillations were not observed in early samples, the first insights came from ARPES studies.
As discussed in Sec.~\ref{sec:ARPES}, SX-ARPES provided the first momentum-resolved Fermi-surface maps over a wide region of reciprocal space, but its limited energy resolution made it difficult to determine the precise Fermi-surface contours~\cite{UTe2_ARPES}.
A later low-$h\nu$ ARPES study with higher energy resolution reported that the Fermi-surface topology is largely described by hypothetical \ThTeTe constructed in the same crystal structure as \UTeTe, corresponding to a localized-$5f$ model of \UTeTe~\cite{UTe2_ARPES_Wray}.
On the other hand, the spectra differ substantially between SX-ARPES and low-$h\nu$ ARPES, suggesting that the surface electronic structure differs significantly from the bulk electronic structure, as discussed in Secs.~\ref{sec:ARPES} and \ref{sec:discussion}.
The low-$h\nu$ ARPES measurements further suggested the presence of a heavy electron pocket around the \pnt{Z} point.
This issue is discussed further in Sec.~\ref{sec:3DFS_calc}.

A major advance was achieved when higher-quality single crystals became available through the molten-salt-flux growth method developed by Sakai \ea~\cite{UTe2_Sakai_MSF}.
Using such improved crystals, Aoki \ea observed de~Haas--van~Alphen (dHvA) oscillations and revealed two orthogonal quasi-one-dimensional Fermi sheets extending along the $k_c$ direction~\cite{UTe2_dHvA}.
This quasi-low-dimensional Fermi-surface topology is further supported by high-field quantum-interference measurements under pressure, which showed smooth evolution up to 19.5~kbar and an increasing frequency interpreted as enhanced warping of the cylindrical Fermi sheets and increased $\mathrm{U}~5f$-orbital weight at \EF~\cite{UTe2_QI_Weinberger_CommPhys2025}.

Figure~\ref{fig:UTe2_dHvA} summarizes the dHvA results for \UTeTe\ and compares them with GGA+$U$ band-structure calculations for \UTeTe and LDA calculations for \ThTeTe.  
The experimentally observed oscillation frequencies [Fig.~\ref{fig:UTe2_dHvA}(a)] are well reproduced by the GGA+$U$ calculation with $U = 2~\mathrm{eV}$ [Fig.~\ref{fig:UTe2_dHvA}(b)], whereas they are not captured by the LDA calculation for \ThTeTe\ [Fig.~\ref{fig:UTe2_dHvA}(c)].
(Details of the differences in the Fermi-surface shapes obtained from these two types of calculations are shown in Fig.~\ref{fig:ThTe2_UTe2} and discussed in Sec.~\ref{sec:TheoryFS}.)
Thus, the dHvA results contradict the low-$h\nu$ ARPES results, although both suggest similar orthogonal quasi-one-dimensional Fermi-surface sheets.

Quasiparticle-interference (QPI) measurements provide information complementary to ARPES and quantum oscillations by probing real-space modulations of the local density of states and the associated scattering wave vectors.
QPI is sensitive to low-energy quasiparticles and surface electronic states, and mainly reflects states projected onto the surface plane rather than resolving the perpendicular momentum component.
Thus, QPI is useful for examining how surface-related states are connected to, or distinguished from, the bulk Fermi surface inferred from bulk-sensitive probes.
A recent QPI study on the (0--11) surface reported superconducting-state Bogoliubov quasiparticle interference governed by the surface-projected normal-state Fermi surface~\cite{UTe2_QPI_Wang}.
The observed QPI wave vectors were reproduced by a model with quasi-one-dimensional Fermi-surface sheets similar to those inferred from dHvA measurements, indicating that QPI can bridge momentum-resolved ARPES spectra and heavy-quasiparticle formation.
Because this study also discussed an odd-parity superconducting surface state, QPI is relevant for clarifying the relationship between the normal-state electronic structure and possible topological superconductivity in \UTeTe.

The Fermi surface determined by ARPES measurements is discussed further in Sec.~\ref{sec:FSmap}.
\subsection{\label{sec:3DFS_calc}3D Fermi surface pocket?}
\begin{figure*}
	\centering
	\includegraphics[scale=0.45]{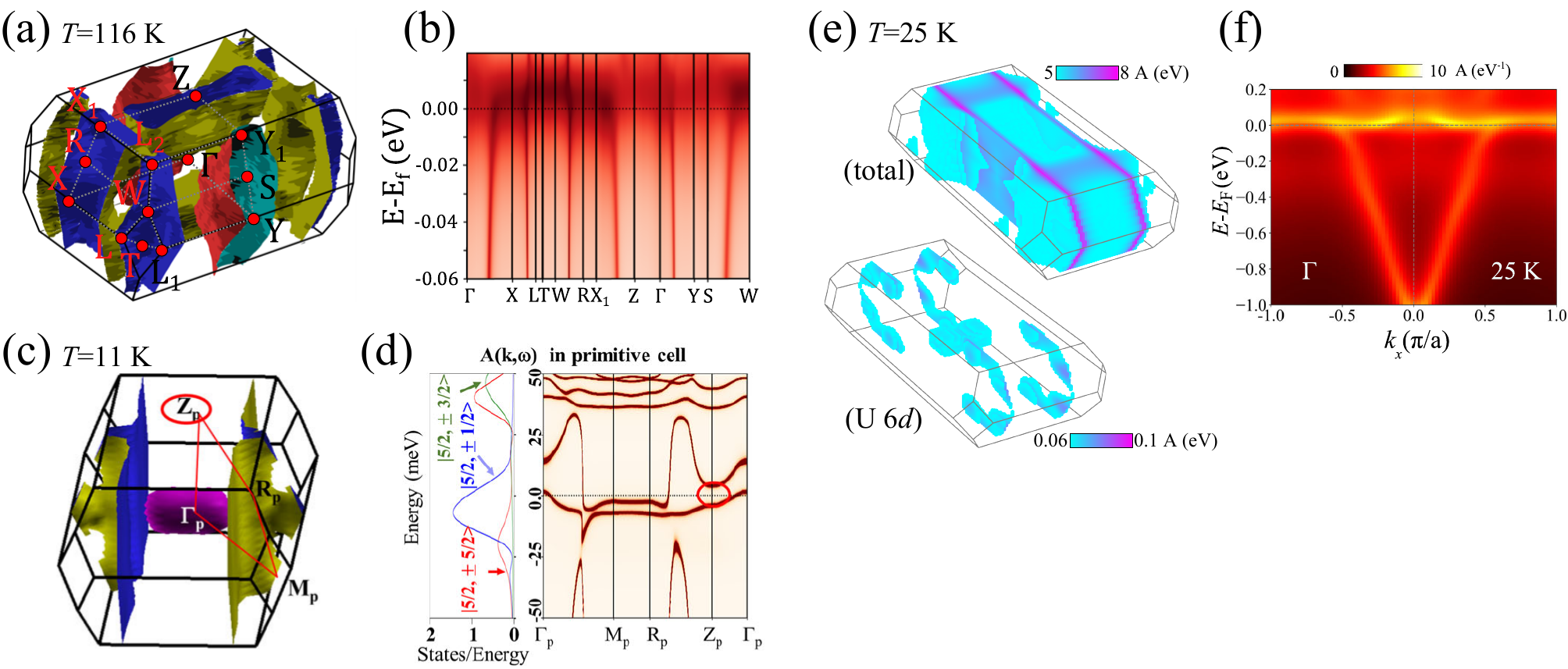}
		\caption{Proposed three-dimensional Fermi surfaces and spectra from DFT+DMFT-based calculations for \UTeTe.
		(a,b) Fermi surface and near-\EF spectral function at $116~\mathrm{K}$ from Halloran \ea, 2025~\cite{DFTDMFT_Halloran}.
		(c,d) Fermi surface and near-\EF quasiparticle spectrum at $11~\mathrm{K}$ from Choi \ea, 2024~\cite{DFTDMFT_Choi}.
		(e,f) Results from Kang \ea: (e) total (upper) and \orb{U}{6d}-projected (lower) Fermi surfaces showing the three dimensional pocket at \Gm~\cite{UTe2_LQSGWDMFT1}; (f) spectral function near \Gm at $25~\mathrm{K}$~\cite{UTe2_LQSGWDMFT2}.
		Panels (a,b), (c,d), and (f) are reproduced from \cite{DFTDMFT_Halloran}, \cite{DFTDMFT_Choi}, and \cite{UTe2_LQSGWDMFT2}, respectively, under the \href{https://creativecommons.org/licenses/by/4.0/}{CC BY 4.0} license and with permission from Springer Nature;
		panel (e) is reprinted from \cite{UTe2_LQSGWDMFT1} with permission, Copyright (2025) by the American Physical Society, .
		}
	\label{fig:UTe2_3DFS}
\end{figure*}
As revealed by de~Haas--van~Alphen (dHvA) measurements, the dominant Fermi surface of \UTeTe is composed primarily of two orthogonal quasi-one-dimensional sheets.  
One of the central unresolved issues concerning its Fermi-surface topology is the possible existence of an additional closed three-dimensional (3D) Fermi-surface pocket.  
The presence or absence of such a pocket has important implications for the superconducting order parameter and for the potential topological nature of superconductivity in \UTeTe~\cite{UTe2_SC_sym}.  

The first experimental indication of a closed 3D Fermi-surface pocket was reported by low-$h\nu$ ARPES measurements, which identified an electron-like band with a band bottom at approximately $300~\mathrm{meV}$ that forms a Fermi surface around the \pnt{Z} point [see Fig.~\ref{fig:UTe2_ARPES_Wray}(g)]~\cite{UTe2_ARPES_Wray}.
Signatures consistent with a similar Fermi surface were also suggested by quantum-oscillation experiments~\cite{UTe2_3DFS}. 
Several recent DFT+DMFT studies have predicted the existence of a closed 3D Fermi-surface pocket~\cite{DFTDMFT_Halloran, DFTDMFT_Choi, UTe2_LQSGWDMFT2, UTe2_LQSGWDMFT1}.  
Figure~\ref{fig:UTe2_3DFS} summarizes representative theoretical results that support this scenario.
Halloran \ea reported that the Fermi surface evolves from a \ThTeTe-like topology at high temperatures to a topology that includes an additional closed 3D Fermi-surface pocket centered at the \pnt{Z} point at low temperatures, as shown in Figs.~\ref{fig:UTe2_3DFS}(a) and (b)~\cite{DFTDMFT_Halloran}.
In contrast, Choi \ea predicted a closed 3D Fermi-surface pocket centered at the $\Gamma$ point, as shown in Figs.~\ref{fig:UTe2_3DFS}(c) and (d)~\cite{DFTDMFT_Choi}.  
Similar closed 3D Fermi-surface features have also been reported in calculations within the GW framework and in combined LQSGW+DMFT approaches, as shown in Figs.~\ref{fig:UTe2_3DFS}(e) and (f)~\cite{UTe2_LQSGWDMFT1,UTe2_LQSGWDMFT2}.  
It should be noted that, in these calculations, the bands responsible for the 3D Fermi-surface pockets are extremely shallow, with their top or bottom located within only a few tens of meV of $E_{\mathrm{F}}$.  
Thus, this energy scale is substantially smaller than the energy dispersion suggested by low-$h\nu$ ARPES ($\sim 300~\mathrm{meV}$), indicating that they are very different in their band structures.

By contrast, other quantum-oscillation studies have concluded that their data can be largely accounted for by two orthogonal quasi-one-dimensional Fermi surfaces, with no compelling evidence for closed 3D pockets~\cite{UTe2_Q2D1,UTe2_Q2D2}.  
Recent high-precision measurements of electronic transport anisotropy are likewise consistent with a predominantly quasi-one-dimensional Fermi surface~\cite{UTe2_eledim}.  

Taken together, these results indicate that the existence of a closed 3D Fermi-surface pocket in \UTeTe remains a matter of active debate.
Because the presence or absence of such a pocket plays a decisive role in identifying the symmetry and topology of the superconducting order parameter, resolving this issue is essential for a comprehensive understanding of the electronic structure of \UTeTe.  
We return to this question from the perspective of ARPES studies in Secs.~\ref{sec:FSmap} and \ref{sec:discussion_ARPES}.

\section{\label{sec:ARPES}ARPES studies of \UTeTe}
ARPES studies of \UTeTe\ were first reported by Fujimori \ea~\cite{UTe2_ARPES}.  
These measurements employed soft X-ray photons with energies in the range $h\nu=565$--$800~\mathrm{eV}$, i.e., SX-ARPES, which are approximately an order of magnitude higher than those used in conventional low-photon-energy ARPES (\hn{\lesssim 100}).  
Subsequently, ARPES experiments using lower photon energies in the range $h\nu=30$--$150~\mathrm{eV}$ (low-$h\nu$ ARPES) were reported~\cite{UTe2_ARPES_Wray,UTe2_ARPES_Wray2}.  
The key experimental conditions of these ARPES studies are summarized in Table~\ref{ARPES_table}.  
In general, SX-ARPES provides enhanced sensitivity to the bulk electronic structure owing to the increased photoelectron mean free path, whereas low-$h\nu$ ARPES offers superior energy resolution and detailed access to near-$E_{\mathrm{F}}$ electronic states.

In the following, we first discuss resonant photoemission spectra to establish the overall energy distribution of the \Uf states in the valence band.  
We then discuss the momentum-resolved band structure revealed by SX-ARPES, followed by a comparison of the near-$E_{\mathrm{F}}$ electronic structure and Fermi surface maps obtained by SX-ARPES and low-$h\nu$ ARPES.  
Finally, the ARPES results for \UTeTe are placed in a broader context through comparison with related uranium-based superconductors, including \UCoGe, \URhGe, and \UGe.

\subsection{\texorpdfstring{Energy distribution of $5f$ states in \UTeTe}{Energy distribution of 5f states in UTe2}}
\begin{figure}
	\centering
	\includegraphics[scale=0.45]{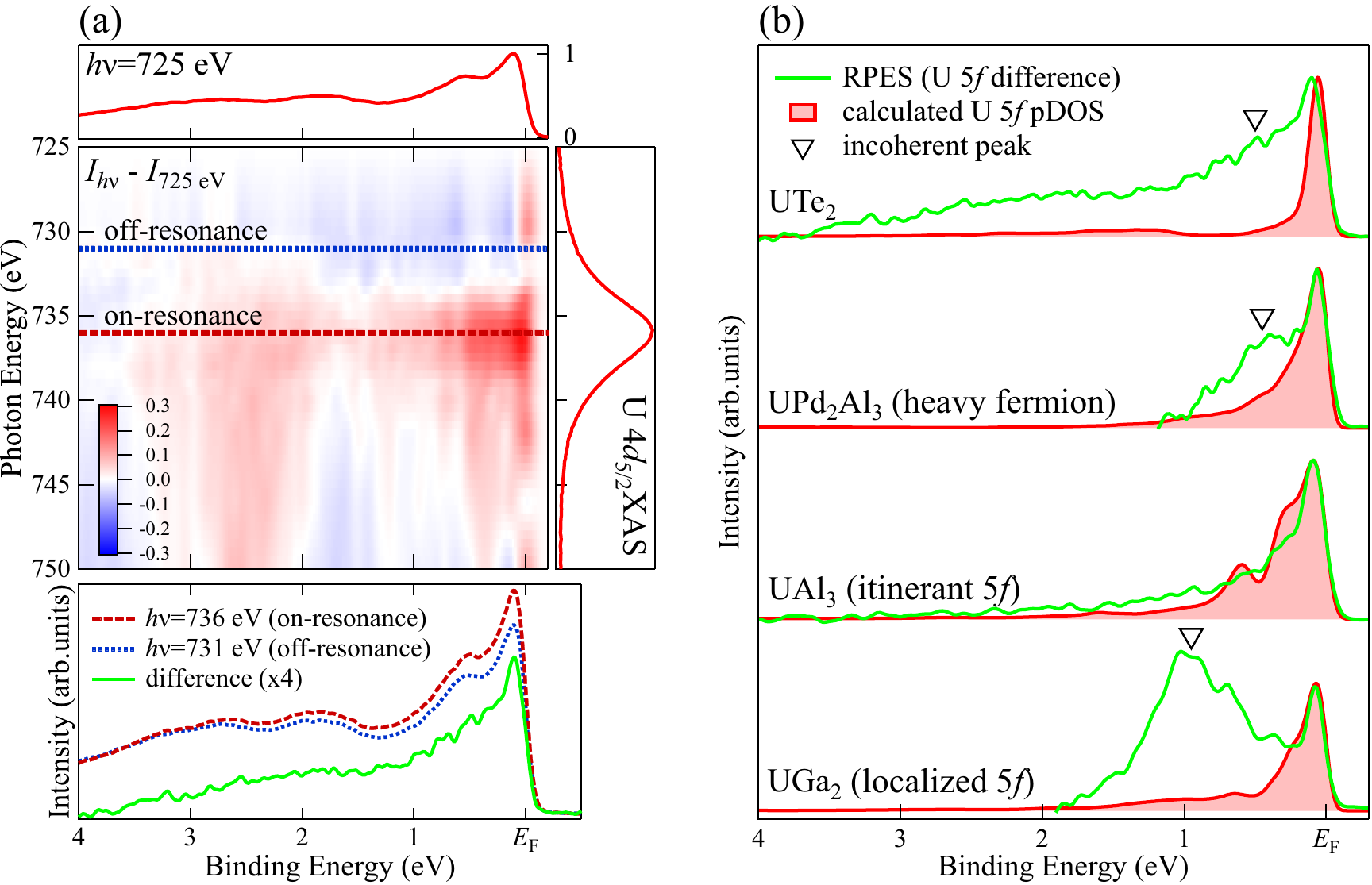}
		\caption{RPES spectra of \UTeTe, together with those of representative uranium compounds.
		(a) Density plot of the RPES spectra of \UTeTe, obtained by subtracting the spectrum measured at \hn{=725} (center), together with the corresponding $\mathrm{U}~4d_{5/2}$ XAS spectrum (right).
		On- and off-resonance spectra were measured at \hn{=736} and $731~\mathrm{eV}$, respectively, and the corresponding difference spectrum is shown at the bottom panel.
		(b) Comparison of the RPES spectra of \UTeTe with those of the heavy-fermion compound \UPdAl, the itinerant-$5f$ compound \UAl, and the localized-$5f$ compound \UGa.
		The corresponding calculated \Uf partial DOS are also shown for comparison.
		Panel (a) is replotted from Fig. 4 of Fujimori \etal, 2019~\cite{UTe2_ARPES}.
		}
l	\label{fig:UTe2_RPES}
\end{figure}
Resonant photoemission (RPES) enhances signals from specific orbitals in photoemission spectra by tuning the incident photon energy to a specific absorption edge.
For uranium compounds, the $5d$ (\hn{\sim 108})~\cite{U5dRPES_Iwan} and $4d$ (\hn{\sim 736})~\cite{U4d5fRPES} absorption edges have been used to enhance the \Uf contributions in the spectra.
However, RPES spectra measured at the $5d$ absorption edge are highly surface sensitive because the kinetic energies of the photoelectrons are near the minimum of the electron inelastic mean free path (IMFP) in solids ($\lesssim 5~\mbox{\AA}$)~\cite{IMFP}.
Thus, we focus here on RPES at the $4d$ absorption edge, which provides sufficient bulk sensitivity ($\gtrsim 15~\mbox{\AA}$).

Figure~\ref{fig:UTe2_RPES} shows the \Udf RPES spectra of \UTeTe, together with those of a typical heavy-fermion compound (\UPdAl), an itinerant compound (\UAl), and a localized compound (\UGa)~\cite{U4d5fRPES}.  
Figure~\ref{fig:UTe2_RPES}(a) presents a density map of the RPES intensity of \UTeTe, where each spectrum is shown after subtraction of the off-resonance spectrum measured at \hn{=725}.  
The corresponding $\mathrm{U}~4d_{5/2}$ XAS spectrum is displayed on the right, with the resonance condition identified around \hn{=736}.  
A pronounced Fano-type enhancement is observed at the $\mathrm{U}~4d_{5/2}$ absorption edge.  
Notably, this enhancement is clearly visible around \EB{\sim 3}, indicating that the \Uf states extend to relatively high binding energies.  
The difference spectrum, obtained by subtracting the off-resonance spectrum (\hn{=731}) from the on-resonance spectrum (\hn{=736}), represents the \Uf-derived contribution and is shown in the bottom panel of Fig.~\ref{fig:UTe2_RPES}(a).  
The extracted \Uf spectral weight exhibits a dominant sharp peak at \EF and extends over an energy range from \EF to approximately \EB{\sim 3}.  

Figure~\ref{fig:UTe2_RPES}(b) compares these RPES-derived \Uf spectrum of \UTeTe with those of \UPdAl, \UAl, and \UGa, together with the calculated \Uf partial densities of states (DOS) of each compound.
For \UAl, the experimental line shape is well reproduced by the calculated DOS, consistent with predominantly itinerant \Uf character.  
In contrast, \UPdAl, \UTeTe, and particularly \UGa exhibit an additional broad feature at higher binding energies, characteristic of an incoherent \Uf component.  
The spectral weight of this incoherent feature in \UTeTe is comparable to that in \UPdAl and significantly smaller than in \UGa, indicating an intermediate degree of electron correlation between the itinerant and localized limits.  
In addition, the \UTeTe spectrum shows a gradual tail extending toward higher binding energies.  
Although its origin remains to be clarified, this feature may be associated with strong hybridization between the \orb{Te}{5p} and \Uf states.

\begin{table*}
\centering
\scalebox{0.75}{
\begin{tblr}{
    hline{1,Z} = {0.1em},
    hline{2} = {0.05em},
    colspec = {c c c c c c},
    cell{1}{1-Z} = {c, font=\bfseries},
    row{1} = {font=\bfseries},
}
Photon energy	& Energy resolution & Temperature & Surface  & Polarization & Reference\\

 565--800~eV  & 90--140~meV	& 20~K & (001) & Unpolarized &  \cite{UTe2_ARPES} \\

30--150~eV &  10--30~meV & 20~K & (001) & Linear ($\pi$ and $\sigma$) & \cite{UTe2_ARPES_Wray} \\

 74~eV, 110~eV & 10--30~meV & 20--110~K & (011) & Linear ($\pi$ and $\sigma$) & \cite{UTe2_ARPES_Wray2} \\
\end{tblr}
}
\caption{Summary of angle-resolved photoemission spectroscopy (ARPES) studies on \UTeTe. 
Experimental conditions such as surface orientation, photon energy range, energy resolution, and polarization are listed for comparison.}
\label{ARPES_table}
\end{table*}
%
\begin{figure*}
	\centering
	\includegraphics[scale=0.45]{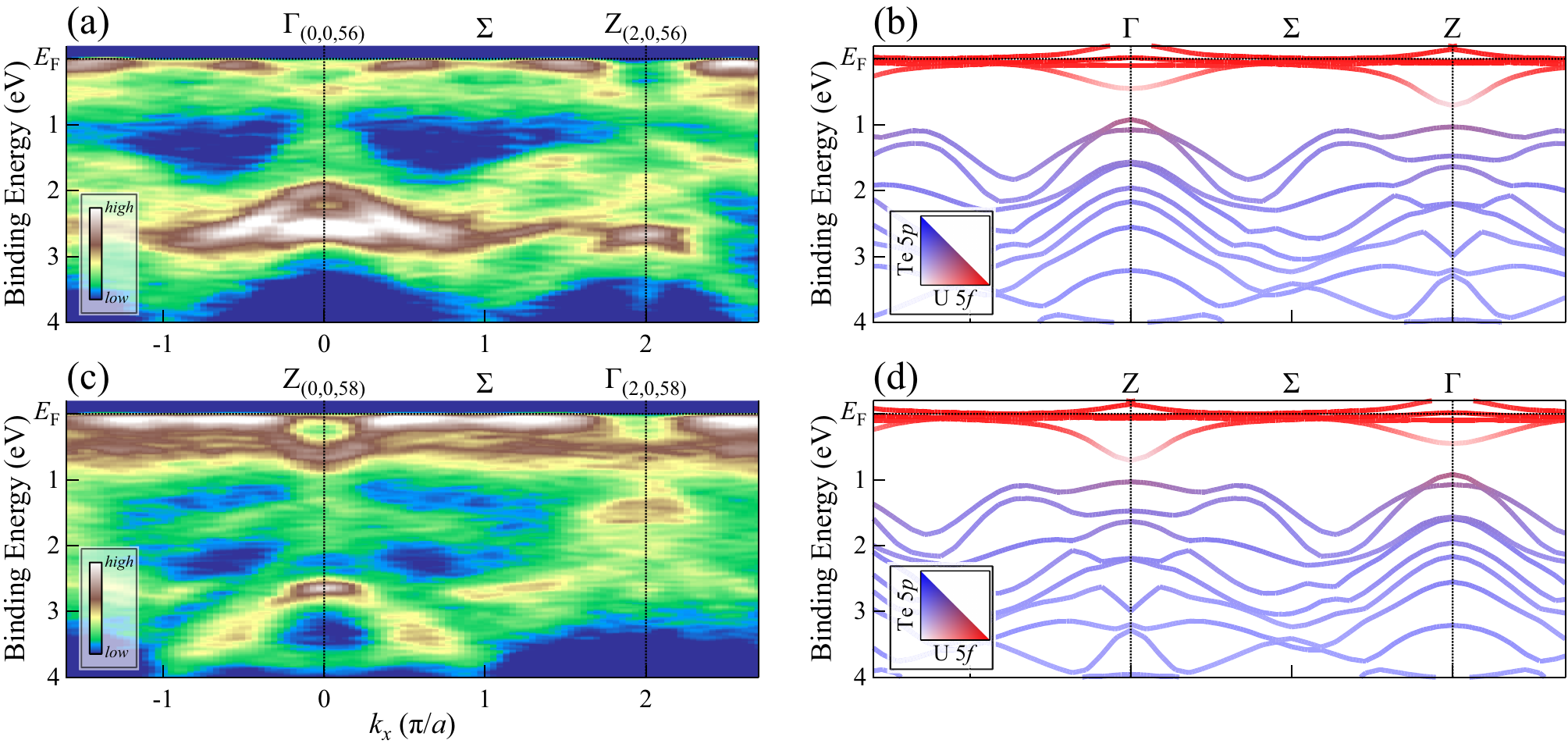}
	\caption{ARPES spectra of \UTeTe measured by soft X-rays.
	(a) ARPES spectra measured at \hn{=580}.
	(b) corresponding band structure calculation.
	(c) APRES spectra measured at \hn{=620}.
	(d) corresponding band structure calculation.
	Reproduced from Fig. 5 of Fujimori \ea, 2019~\cite{UTe2_ARPES}.
}
	\label{fig:UTe2_ARPES_SX}
\end{figure*}

\subsection{Band structure}
To understand the overall band structure of \UTeTe, we first present the ARPES spectra over a wide energy range.
Figure~\ref{fig:UTe2_ARPES_SX} compares the SX-ARPES spectra with the calculated band structure~\cite{UTe2_ARPES}.
Figure~\ref{fig:UTe2_ARPES_SX}(a) shows the ARPES spectra of \UTeTe measured along the $\mathrm{\Gamma}_{(0,0,56)}$--$\mathrm{(\Sigma)}$--$\mathrm{X}_{(2,0,56)}$ high-symmetry line.
The coordinates (0,0,56) and (2,0,56) denote positions in momentum space, where the units are $\pi/a = 0.7550~[\mbox{\AA}^{-1}]$, $\pi/b =0.5131~[\mbox{\AA}^{-1}]$, and $\pi/c = 0.2251~[\mbox{\AA}^{-1}]$ for the $k_x$, $k_y$, and $k_z$ directions, respectively.
These spectra were obtained with \hn{\sim 580} along the ARPES cut indicated in the multi-BZ momentum map in Fig.~\ref{fig:UTe2_BZ}(b).
Weakly dispersive features near \EF appear as quasiparticle (QP) bands mainly derived from the \Uf states.
Their intensities depend on momentum close to \EF, indicating that they form Fermi surfaces.
At higher binding energies (\EB{\gtrsim 1}), more dispersive bands appear.
These bands originate mainly from the \orb{Te}{5p} states, although a small contribution from the \Uf states is also present, as indicated by the RPES results.

Figure~\ref{fig:UTe2_ARPES_SX}(b) shows the corresponding calculated band structure.
The color represents the relative weight of the \Uf and \orb{Te}{5p} states.
In general, the bands at \EB{\lesssim 1} and \EB{\gtrsim 1} are dominated by the \Uf and \orb{Te}{5p} states, respectively.
Although individual bands are not well resolved in the experimental spectra, the overall features agree well with the calculation.
For example, the inverted parabolic dispersions near the \Gm and \pnt{X} points at \EB{\gtrsim 1} correspond to the calculated bands at similar energies.
The weakly dispersive $5f$-originated features just below \EF are also consistent, within the experimental energy resolution ($\sim 80~\mathrm{meV}$), with the calculated \Uf-derived bands, as discussed in more detail in Sec.~\ref{sec:nearEF}.

Figures~\ref{fig:UTe2_ARPES_SX}(c) and (d) show the ARPES spectra measured along the $\mathrm{Z}_{(0,0,58)}$--$\mathrm{(\Sigma)}$--$\mathrm{\Gamma}_{(2,0,58)}$ line and the corresponding calculation, respectively.
These spectra were measured with \hn{\sim 620} along the cut shown in Fig.~\ref{fig:UTe2_BZ}(b).
The main structure of the spectra is similar to that along the $\mathrm{\Gamma}_{(0,0,56)}$--$\mathrm{(\Sigma)}$--$\mathrm{Z}_{(2,0,56)}$ line, but the relative intensities differ due to matrix-element effects.
These differences help trace the energy dispersions more clearly.
In particular, the intensity around the \pnt{Z} point is enhanced.
As for the other cut, the overall structure of the ARPES spectra is well reproduced by the calculated bands.

\subsection{\label{sec:nearEF}Electronic structure near \texorpdfstring{$E_{\mathrm{F}}$}{EF}}
\begin{figure*}
	\centering
	\includegraphics[scale=0.44]{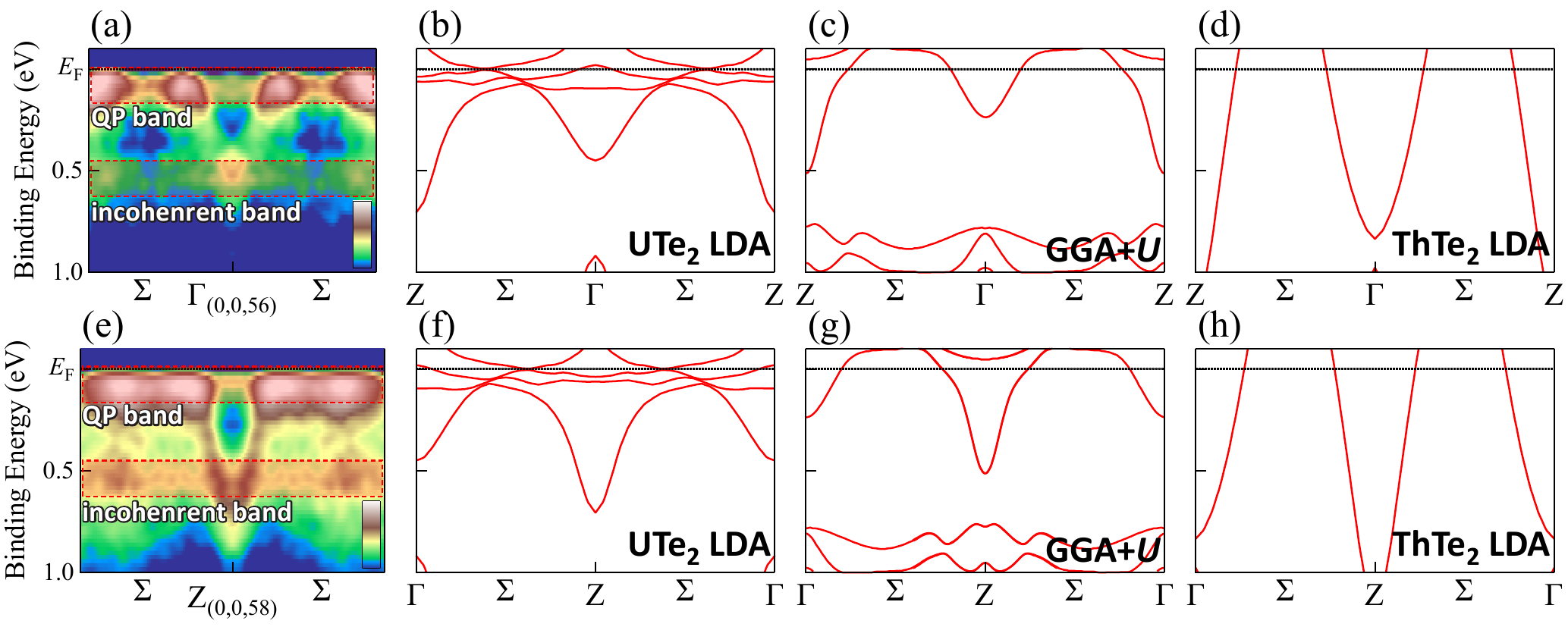}
	\caption{Comparison of the near-\EF electronic structure of \UTeTe with various electronic structure calculations.
	(a) near-\EF ARPES spectra around \Gm point measured at \hn{=580}.
	(b) LDA for \UTeTe.
	(c) GGA+$U$ with $U=2~\mathrm{eV}$.
	(d) LDA for \ThTeTe, which corresponds to the localized-\Uf limit.
	(e) near-\EF ARPES spectra around \pnt{X} point measured at \hn{=620}.
	(f) LDA for \UTeTe.
	(g) GGA+$U$ with $U=2~\mathrm{eV}$.
	(h) LDA for \ThTeTe, which corresponds to the localized-\Uf limit.
}
	\label{fig:UTe2_ARPES_EF}
\end{figure*}
%
Figure~\ref{fig:UTe2_ARPES_EF} compares the near-\EF electronic structure of \UTeTe with various band-structure calculations.  
Figure~\ref{fig:UTe2_ARPES_EF}(a) shows the ARPES spectra measured at \hn{=580}.  
An electron-like band centered at the $\Gamma$ point is clearly observed.  
Its spectral weight becomes strongly enhanced and shows a pronounced renormalization near \EF, evidencing the formation of heavy quasiparticles.  
In addition, a nearly flat incoherent feature is also present around \EB{\sim 0.6}.

Figures~\ref{fig:UTe2_ARPES_EF}(b)--(d) display three representative calculations:  
(b) LDA for \UTeTe,  
(c) GGA+$U$ with $U=2$~eV, and  
(d) LDA for \ThTeTe, representing the localized-\Uf limit.  
In the LDA calculation for \UTeTe, the \Uf spectral weight remains concentrated near \EF, whereas in the GGA+$U$ calculation it is pushed to higher binding energies.  
Notably, all three calculations predict an electron-like band at the $\Gamma$ point, although the corresponding band-bottom energies differ substantially.

We now compare the experimental ARPES spectra with the calculated band structures shown in Figs.~\ref{fig:UTe2_ARPES_EF}(b)--(d) and (f)--(h).
The experimentally observed incoherent feature is absent in the LDA calculation.
Although the GGA+$U$ calculation reproduces the overall Fermi-surface topology, it does not capture the strong low-energy renormalization: the calculated electron-like bands show little mass enhancement near \EF, and the incoherent (lower-Hubbard-band) feature appears at a higher binding energy than observed (\EB{\sim 0.9}).
These discrepancies indicate that more advanced theoretical methods are required to describe the correlated \Uf electrons in \UTeTe.
The LDA calculation for \ThTeTe also yields an electron-like band at $\Gamma$, but its near-\EF structure differs substantially from experiment, and the band bottom is located much deeper in energy.
This mismatch shows that a nearly localized $5f^2$ description alone is incompatible with the observed electronic structure of \UTeTe.
A similar discrepancy is found near the \pnt{Z} point, as shown in Figs.~\ref{fig:UTe2_ARPES_EF}(e)--(h).
It is also noteworthy that the experimental band-bottom energy of the electron-like band at the $\Gamma$ point is closest to that obtained from the LDA calculation for \UTeTe, suggesting that treating the \Uf states as at least partially itinerant is essential for reproducing the observed band structure.

\subsection{\label{sec:FSmap}Fermi surface map}
\begin{figure}
	\centering
	\includegraphics[scale=0.45]{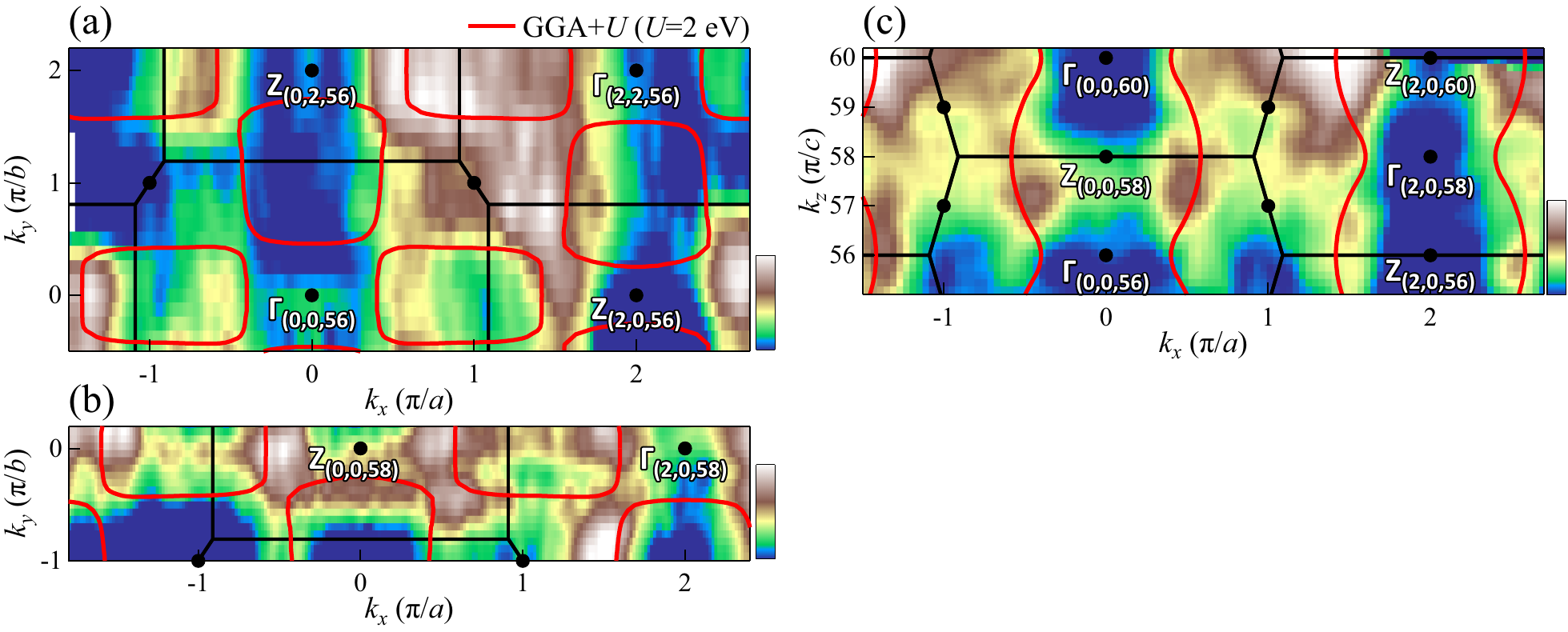}
	\caption{Fermi surface map of \UTeTe and the calculated Fermi surface by GGA+$U$ ($U=2~\mathrm{eV}$.
	(a) Fermi surface map obtained by $h\nu$ scan at \hn{=565-625}.
	(b) Fermi surface map measured at \hn{=580}.
	(c) Fermi surface map measured at \hn{=620}.
}
	\label{fig:UTe2_SX_EFmap}
\end{figure}
%
Figure~\ref{fig:UTe2_SX_EFmap} shows the Fermi-surface map of \UTeTe obtained by integrating the photoemission intensity within $\pm50~\mathrm{meV}$ of \EF.
Figures~\ref{fig:UTe2_SX_EFmap}(a) and (b) show Fermi-surface maps  in the $k_x$--$k_y$ ($k_a$--$k_b$) plane around the \Gm point (measured at \hn{=580}) and around the X point (measured at \hn{=620}), respectively.  
Figure~\ref{fig:UTe2_SX_EFmap}(c) presents the Fermi-surface map in the $k_x$--$k_z$ ($k_a$--$k_c$) plane, obtained by scanning the photon energies from \hn{=565} to \hn{=625}.
The corresponding ARPES cut is indicated in Fig.~\ref{fig:UTe2_BZ}(b).
The calculated Fermi surface from the GGA+$U$ calculation ($U=2~\mathrm{eV}$) is overlaid as solid curves.
It should be noted that the photoemission intensities vary across different regions of the Brillouin zone, even where the underlying symmetry is equivalent.
This variation arises from photoemission matrix-element effects, which can strongly modulate the observed intensity~\cite{SF_review_JPCM, SF_review_JPSJ}.
Therefore, measuring the same high-symmetry points with different photon energies and detection geometries is essential for reliably determining the presence or absence of Fermi-surface features.

The agreement between experiment and theory is not fully quantitative, partly because the heavy bands just below \EF, contribute finite intensity to the Fermi-surface maps.
Because the \Uf states form very narrow bands near \EF, the observed intensity contains contributions not only from the states forming the Fermi surface but also from flat bands lying just below \EF.
Nevertheless, the overall intensity distribution exhibits a pronounced one-dimensional character, as the spectral weight along the \Gm--\pnt{Z}--\Gm high-symmetry direction is relatively weak.
Notably, only weak intensities are observed around the \Gm and \pnt{Z} points, suggesting the absence of heavy Fermi-surface pockets at these locations.
Overall, the Fermi-surface topology and its partially low-dimensional character revealed by ARPES are consistent with the highly anisotropic normal-state and superconducting properties of \UTeTe, implying that the itinerant \Uf electrons play a key role in the formation of its unconventional superconducting state.

\subsection{\texorpdfstring{Low-$h\nu$ ARPES of \UTeTe}{Low-hnu ARPES of UTe2}}
\begin{figure*}
	\centering
	\includegraphics[scale=0.45]{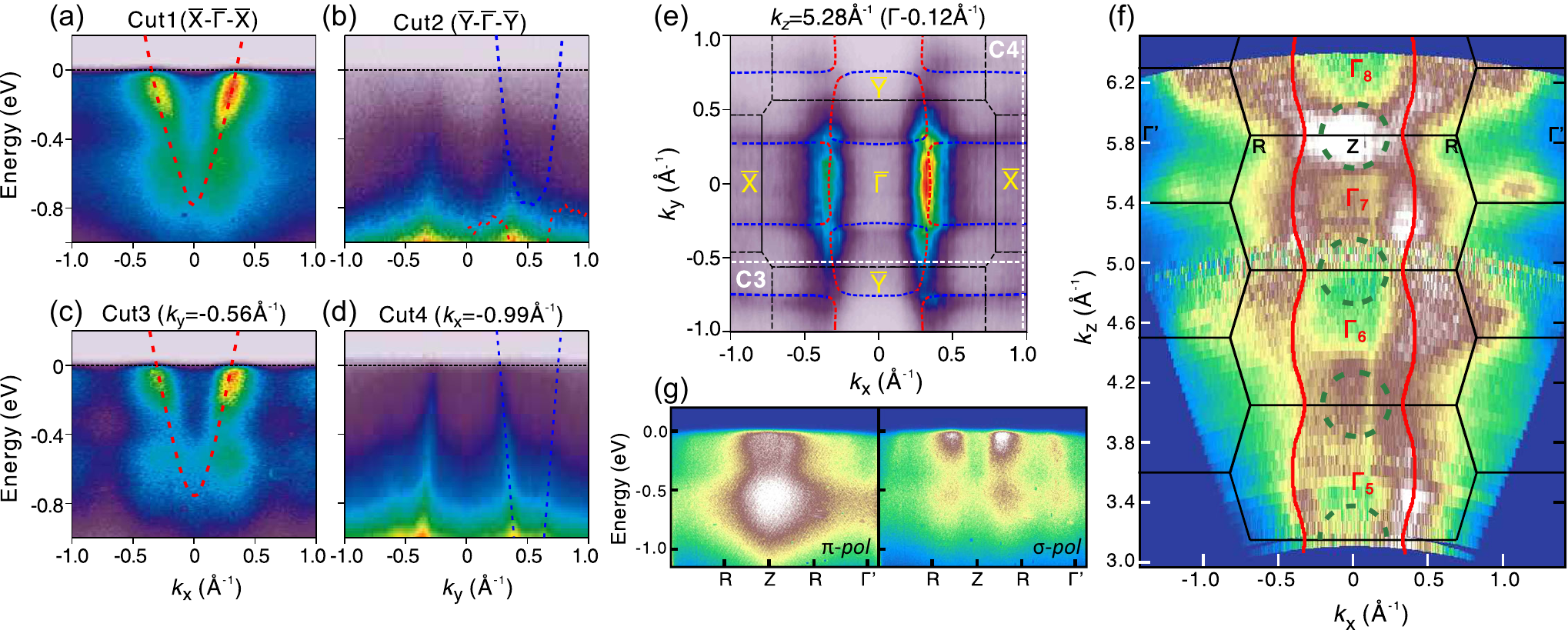}
	\caption{ARPES spectra and Fermi surface maps of \UTeTe obtained by low-$h\nu$ ARPES.
	(a,b) ARPES spectra measured at \Uedge{O} resonance (\hn{=98}) along $k_x$ ($k_a$) and $k_y$ ($k_b$) directions, respectively.
	(c, d) ARPES spectra measured at off resonance (\hn{=92}) along Cut3 and Cut4 directions, respectively.
	See Fig.~\ref{fig:UTe2_ARPES_Wray} (e) for the location of ARPES cut.
	(e) Fermi surface map measured at \Uedge{O_{4,5}} resonance (\hn{=98}).
	The Fermi surfaces calculated for \ThTeTe, representing the localized $5f^2$ configuration and shifted upward by $200~\mathrm{meV}$, are superimposed by dotted curves.
	(f) Fermi surface map, measured by tuning photon energy from $h\nu=30$ to 150~eV.
	The calculated Fermi surface of \ThTeTe, plotted as solid curves, is the same as that used for the in-plane comparison. 
	(g) Polarization dependent ARPES spectra of the \pnt{Z} point measured at \hn{=125}, showing (left) $\pi$ polarization and (right) $\sigma$ polarization.
	Reprinted from\cite{UTe2_ARPES_Wray} with permission, Copyright (2020) by the American Physical Society.
}
	\label{fig:UTe2_ARPES_Wray}
\end{figure*}
Complementary low-$h\nu$ ARPES measurements ($h\nu = 30$--$150~\mathrm{eV}$) have also been reported~\cite{UTe2_ARPES_Wray}.  
Figures~\ref{fig:UTe2_ARPES_Wray}(a) and (b) show representative spectra measured at \hn{=98} along the $k_x$ ($k_a$) and $k_y$ ($k_b$) directions, respectively.
The corresponding ARPES cut is indicated in Fig.~\ref{fig:UTe2_BZ}(b).
To guide the comparison, the calculated band dispersions for \ThTeTe, which is used as a localized-$5f^{2}$ reference for \UTeTe and shifted upward by $200~\mathrm{meV}$, are overlaid as dotted curves.  
Along $k_x$ [Fig.~\ref{fig:UTe2_ARPES_Wray}(a)], an electron-like light band is observed around the \Gm point and is reproduced well by the calculation.  
Along the orthogonal high-symmetry line $k_y$ [Fig.~\ref{fig:UTe2_ARPES_Wray}(b)], hole-like light bands are similarly observed and remain consistent with the same calculation.  
A comparable trend is found for Cuts 3 and 4 [Figs.~\ref{fig:UTe2_ARPES_Wray}(c)--(d)], where only light, fast-dispersing bands are detected.

At first glance, these dispersions resemble parts of the SX-ARPES results; however, an important difference appears near \EF.  
For example, the $k_x$ cut in Fig.~\ref{fig:UTe2_ARPES_Wray}(a), which follows nearly the same symmetry line as the \hn{=580} data in Fig.~\ref{fig:UTe2_ARPES_EF}(a), shows no heavy quasiparticle features near \EF.

Figures~\ref{fig:UTe2_ARPES_Wray}(e) and (f) present the corresponding Fermi-surface maps from low-$h\nu$ ARPES.  
The in-plane map at \hn{=98} [Fig.~\ref{fig:UTe2_ARPES_Wray}(e)] is compared with the calculated Fermi surface of \ThTeTe.  
The overall agreement suggests that the in-plane Fermi-surface topology is reasonably captured within a localized-$5f$ description, in marked contrast to the SX-ARPES Fermi-surface maps shown in Figs.~\ref{fig:UTe2_SX_EFmap}(a) and (b).  
Figure~\ref{fig:UTe2_ARPES_Wray}(f) shows the out-of-plane Fermi-surface map obtained by varying the photon energy over $h\nu = 30$--$150~\mathrm{eV}$, together with the same calculated Fermi surface.  
In this geometry, however, the spectra contain a substantial and nonuniform background from incoherent processes, which obscures the Fermi crossings.  
Consequently, the correspondence between experiment and calculation is less clear than in the in-plane case.

Overall, SX-ARPES and low-$h\nu$ ARPES yield markedly different spectra.
These discrepancies are mainly due to differences in bulk sensitivity, as discussed in detail in Sec.~\ref{sec:discussion}.

\subsection[ARPES view of electronic states near the Z point]{ARPES view of electronic states near the \pnt{Z} point}
One notable feature reported in low-$h\nu$ ARPES is a possible closed 3D Fermi-surface pocket at the \pnt{Z} point.
As shown in Fig.~\ref{fig:UTe2_ARPES_Wray}(f), an enhancement of spectral intensity is observed in the vicinity of \pnt{Z}, between the \pnt{\Gamma_7} and \pnt{\Gamma_8} points.
Figure~\ref{fig:UTe2_ARPES_Wray} (g) presents ARPES spectra measured around the \pnt{Z} point at \hn{=125}.
The $\pi$-polarized spectra (left panel) exhibit a broad spectral feature near the \pnt{Z} point close to \EF, with spectral weight extending down to a binding energy of approximately $E_{\mathrm{B}} \simeq 300~\mathrm{meV}$, whereas the $\sigma$-polarized spectra (right panel) display a clear electron-like dispersion associated with the quasi-one-dimensional Fermi surface.
The feature below $400~\mathrm{meV}$ corresponds to the contribution from the incoherent component.
The broad feature observed in the $\pi$-polarized spectra has been discussed as being compatible with a shallow and weakly dispersive heavy electronic state, potentially forming a closed three-dimensional Fermi-surface pocket at the \pnt{Z} point.
However, its dispersion cannot be resolved unambiguously because of the unusually large intrinsic linewidth.

In contrast, no corresponding feature has been identified in SX-ARPES measurements, which provide enhanced bulk sensitivity.  
The Fermi-surface maps obtained by SX-ARPES with various in-plane and out-of-plane momentum scans, shown in Fig.~\ref{fig:UTe2_ARPES_EF} (e), do not reveal any intensity enhancement near the \pnt{Z} point.
Consistently, SX-ARPES spectra measured in the same momentum region show no discernible signature of a corresponding state, as illustrated in Fig.~\ref{fig:UTe2_ARPES_EF}(e).
Given that SX-ARPES probes a substantially larger photoelectron escape depth, the absence of this feature suggests that the intensity enhancement observed near \pnt{Z} point in low-$h\nu$ ARPES may not originate from a coherent bulk band, but could instead reflect surface-related electronic states.

We return to this issue, and its implications for the DMFT-predicted 3D Fermi-surface pockets in \UTeTe, in Sec.~\ref{sec:discussion}.

\subsection{\label{sec:UTe2_relat}\UTeTe and related compounds}
\begin{figure*}
	\centering
	\includegraphics[scale=0.44]{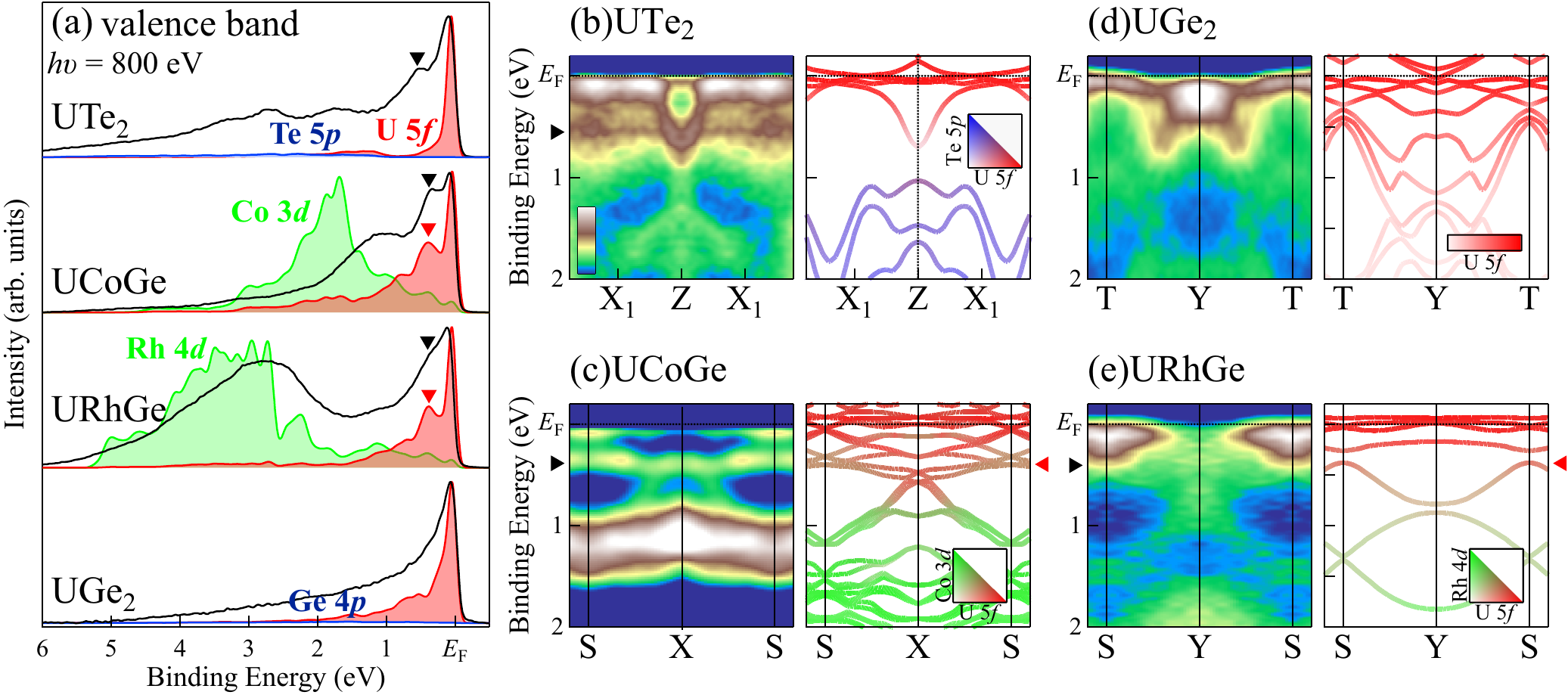}
	\caption{Comparison of \UTeTe and uranium ferromagnetic superconductors.
	(a) AIPES spectra of \UTeTe, together with those of \UCoGe, \URhGe, and \UGe measured at \hn{=800}.
	The calculated \Uf and ligand $d$ and $p$ DOS are also shown.
	The locations of shoulder structures are indicated by black and red inverted triangles.
	(b--e) Representative ARPES spectra of these compounds.
	The locations of the shoulder structures in panel (a) are also indicated by triangles.
	Panels (d-e) are reprlotted from Fig. 26 of Fujimori \etal, 2016~\cite{SF_review_JPSJ}
}
	\label{fig:UTe2_UFM}
\end{figure*}
Because \UTeTe is frequently discussed in relation to the ferromagnetic superconductors \UCoGe, \URhGe, and \UGe, comparing their valence-band electronic structures provides useful context.
Figure~\ref{fig:UTe2_UFM}(a) shows valence-band spectra measured at \hn{=800}, together with the corresponding calculated density of states (DOS) for the \Uf and ligand $p$ and $d$ states \cite{UTe2_ARPES,URhGe_ARPES,UGe2_UCoGe_ARPES}.
Photoionization cross sections for each orbital are multiplied to simulate the spectra from the calculated DOS.
At this photon energy, the photoionization cross sections of the \Uf and transition-metal (TM) $d$ states are enhanced, enabling a direct comparison of the \Uf and transition metal-$d$ contributions.

All spectra exhibit a sharp peak at \EF, consistent with the calculated \Uf DOS.
In addition, the spectra of \UTeTe, \UCoGe, and \URhGe show a common shoulder-like feature at \EB{\sim 0.4-0.6}, as indicated by black inverted triangles.
However, the microscopic origin of this feature differs between \UTeTe and \UCoGe/\URhGe.
In \UCoGe and \URhGe, the shoulder is well reproduced by the calculated \Uf DOS, with its energy position indicated by red inverted triangles, whereas  in \UTeTe it is not captured by the calculation.

This contrast can be understood from the ARPES spectra.
Figures~\ref{fig:UTe2_UFM}(b)--(e) show representative ARPES spectra for these compounds.
In these figures, the locations of the shoulder structures in Fig.~\ref{fig:UTe2_UFM}(a) are also indicated by triangles.
Overall, the ARPES spectra of \UCoGe and \URhGe are largely consistent with the band-structure calculations and exhibit nearly flat bands at \EB{\sim 0.4} that correspond closely to the calculated dispersions.
By contrast, the ARPES spectra of \UTeTe shows an additional incoherent feature around \EB{\sim 0.6} that is not reproduced by the calculation.
This comparison indicates that the prominent structures in \UCoGe and \URhGe predominantly reflect coherent band states, and it suggests that correlation effects in these compounds are comparatively weaker than in \UTeTe.

A qualitatively different behavior is observed in \UGe, whose valence-band spectrum displays a pronounced high-binding-energy tail relative to the calculated DOS.
A similar enhancement is also found in the resonant photoemission spectrum of \UTeTe, suggesting a possible common underlying mechanism.
Such high-binding-energy spectral weight may reflect correlation-enhanced hybridization and an increased admixture of ligand $p$ states into the \Uf manifold.
The absence of comparable features in the resonant photoemission spectrum of the more itinerant compound \UAl further underscores the distinctive role of correlation-enhanced hybridization in \UTeTe and \UGe.

Taken together, this comparative ARPES view suggests that the \Uf states in these uranium superconductors are broadly itinerant, but that \UTeTe exhibits stronger correlation effects than \UCoGe, \URhGe, and \UGe.

\section{\label{sec:Xray}X-ray spectroscopy for \texorpdfstring{$\mathrm{UTe}_2$}{UTe2}}
\begin{table*}
\centering
\scalebox{0.75}{
\begin{tblr}{
    hline{1,Z} = {0.1em},
    hline{2} = {0.05em},
    colspec = {l l l c },
    cell{1}{1-Z} = {c, font=\bfseries},
    row{1} = {font=\bfseries},
}
Probe(s)								& Photon Energy		& Results														& Reference\\
\Uedge{L_3} XANES				& $\sim 17	$~keV	& Intermediate valence, closer to \Unf{2}				& \cite{UTe2_L3_XANES}\\
\Uff core-level						& 800~eV				& Intermediate valence, closer to \Unf{3}				& \cite{UTe2_core}\\
\Uedge{O} XAS, RIXS				& $\sim 100$~eV	& $\mathrm{U}~5f^{2}6d^{1}$-based configuration& \cite{UTe2_O_RIXS}\\
\Uedge{M_{4,5}} XANES(sum rule), XMCD	& $3.5-3.7$~keV	& Intermediate valence (\nf{=2.6-2.8})		& \cite{UTe2_XMCD}\\
\Uedge{M_{4,5}} RIXS			& $3.5-3.7$~keV	& $\mathrm{U}~5f^{2}6d^{1}$-based configuration& \cite{UTe2_M_RIXS}	\\
\Uedge{L_3} XANES, RXES		& $\sim 17$~keV	& Intermediate valence (\nf{\sim 2.26})				& \cite{UTe2_RXES}\\
\Uff core-level (DFT+DMFT)		& 6~keV				& Intermediate valence (\nf{\sim 2.5})					& \cite{UTe2_U4fDMFT} \\
\end{tblr}
}
\caption{Summary of X-ray spectroscopy studies on \UTeTe.  
Each study employed different X-ray spectroscopic techniques to probe the uranium $5f$ electronic configuration.}
\label{xrays_table}
\end{table*}

\begin{figure*}[t]
	\centering
	\includegraphics[scale=0.45]{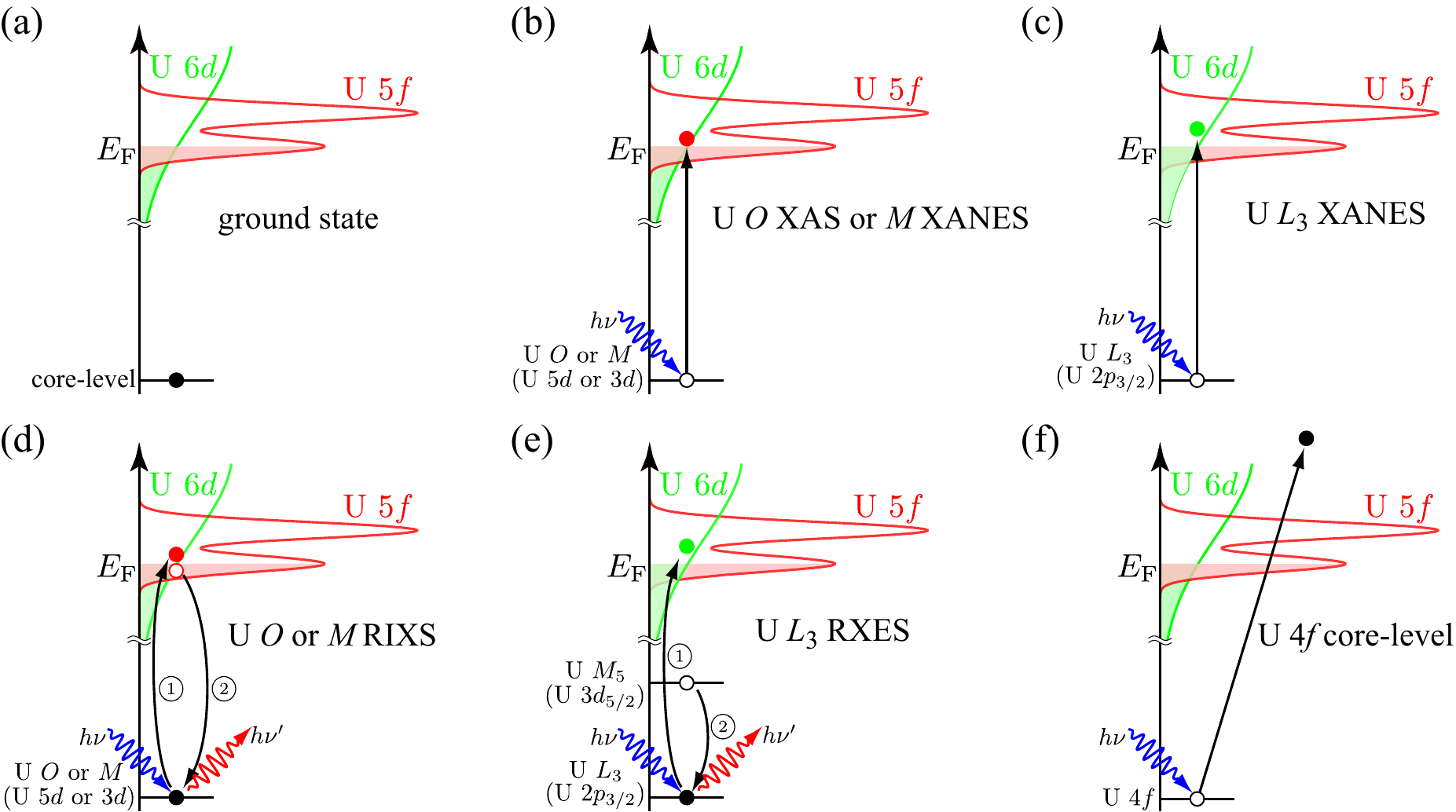}
	\caption{Schematic diagram of X-ray spectroscopies for \UTeTe
	(a) the ground state
	(b) \Uedge{M} or $O$ XANES
	(c) \Uedge{L_3} XANES
	(d) \Uedge{M} or $O$ RIXS
	(e) \Uedge{L} RXRS
	(f) \orb{U}{4f} core-level XPS
}
	\label{xrays_fig}
\end{figure*}
Several X-ray spectroscopy studies have been performed on \UTeTe since the discovery of its superconductivity, and here we summarize the techniques that have been applied.  
Table~\ref{xrays_table} lists the X-ray spectroscopy studies reported for \UTeTe, while Fig.~\ref{xrays_fig} schematically illustrates the corresponding experimental methods.  
These techniques involve specific core-electron excitations, as shown in Fig.~\ref{xrays_fig}(a).  
The details of the applications of these techniques for actinide materials are described in \cite{SR_Actinides,SR_Actinides2}.

In general, X-ray spectroscopy provides an element-specific probe of the local electronic structure around the absorbing atom and can be sensitive to its valence state, orbital occupation, and hybridization with ligand states.
Because X-ray absorption is governed by electric-dipole selection rules, the information obtained depends strongly on the core level being excited.
For uranium compounds, absorption at edges involving $d \rightarrow f$ transitions, such as the $O$, $N$, and $M$ edges, directly probes unoccupied \Uf states.
By contrast, absorption at the $L$ edge corresponds primarily to a $p \rightarrow d$ excitation and therefore probes the unoccupied \orb{U}{6d} states; information on the \Uf states is obtained only indirectly through their hybridization with the $6d$ and ligand orbitals.

\subsection{XAS and XANES}
In X-ray absorption spectroscopy (XAS), core electrons are excited into unoccupied electronic states by incident X-rays.  
The spectral features near the absorption threshold, referred to as X-ray absorption near-edge structure (XANES), primarily probe the unoccupied density of states and are therefore sensitive to the valence configuration.  
In strongly correlated materials, electron--electron interactions can substantially modify the spectra, producing complex satellite structures.

For uranium compounds, XAS (or XANES; Fig.~\ref{xrays_fig}(b)) at the $O$ ($5d$) edge ($h\nu \sim 100~\mathrm{eV}$), the $N$ ($4d$) edges ($h\nu \sim 778$ and $736~\mathrm{eV}$ for the $N_4$ and $N_5$ edges, respectively), and the $M$ ($3d$) edges ($h\nu \sim 3728$ and $3552~\mathrm{eV}$ for the $M_4$ and $M_5$ edges, respectively) probe direct excitations from these core levels into unoccupied \Uf states.
Among these, it has been pointed out that direct $d \rightarrow f$ XAS (at the $O$, $N$, and $M$ edges) shows only weak sensitivity to the uranium valence state.

For the $O$ and $N$ edges, Tobin \ea demonstrated the apparent lack of chemical sensitivity in the corresponding XAS spectra~\cite{U_XAS_Tobin}.
For the $M$ edges, high-energy-resolution fluorescence-detected X-ray absorption (HERFD-XAS), which significantly reduces the lifetime broadening in conventional XAS, has been applied to actinide materials~\cite{AC_HERFD_reivew}.
This method is valuable for estimating the valence state in ionic materials from the chemical shift of the main line.
By contrast, for metallic compounds the chemical shift is very small and the spectra are largely featureless and do not show significant changes from compound to compound~\cite{AC_M_RIXS,UC_HERFD}.
This point is further demonstrated in Appendix~A.
Nevertheless, the XAS branching ratio can still provide information on the $5f$ electron count $n_{5f}$ via the spin--orbit sum rule~\cite{SOsumrule}.

In contrast, the \Uedge{L_3}-edge XANES (\hnk{\sim 17}; Fig.~\ref{xrays_fig}(c)) involves transitions from the \Uedge{L_3} core level into unoccupied \orb{U}{6d} states, thereby probing the \Uf shell indirectly via $5f$--$6d$ interactions.
The \Uedge{L_3}-edge energy position is generally sensitive to the uranium valence and is widely used to estimate $n_{5f}$ for ionic compounds, in which the $6d$ orbitals are unoccupied.
However, for intermetallic compounds, where partially occupied $6d$ bands contribute to core-hole screening, the simple chemical-shift trends known from ionic systems do not apply~\cite{SR_Actinides}.
To estimate the absolute uranium valence from \Uedge{L_3} spectra of intermetallic uranium compounds, it is necessary to model the unoccupied density of states.

Meanwhile, X-ray magnetic circular dichroism (XMCD), which measures the difference in absorption of left- and right-circularly polarized X-rays in the presence of a magnetic field, is a unique tool to probe the orbital ($\mu_L$) and spin ($\mu_S$) magnetic moments carried by the $5f$ electrons.
By comparing the total $5f$ magnetic moment ($\mu_{\mathrm{total}}=\mu_L+\mu_S$) and measured magnetization, the $5f$ count can be estimated.

In this subsection, we summarize XAS and XANES measurements performed on \UTeTe.

\subsubsection{\Uedge{O}-edge XAS}
\begin{figure*}
	\centering
	\includegraphics[scale=0.45]{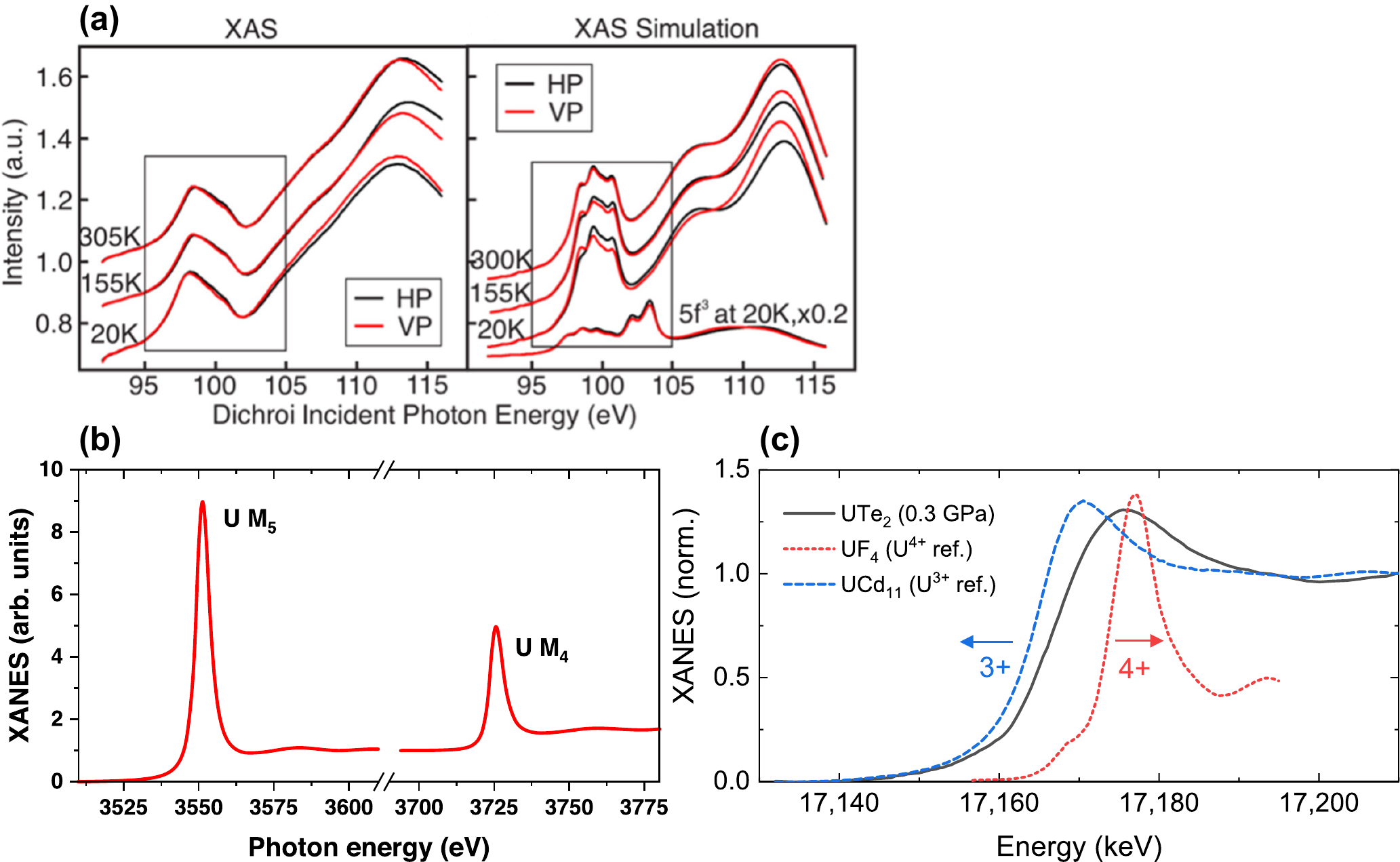}
		\caption[XANES/XAS spectra of UTe2 at O, M4,5, and L3 edges]{XANES (XAS) spectra of \UTeTe measured at various uranium absorption edges:
		(a) O-edge XAS measured at different temperatures with horizontal (HP) and vertical (VP) incident photon polarizations, with multiplet simulations for the $5f^{2}$ and $5f^{3}$ configurations;
		(b) $M_5$- and $M_4$-edge XANES spectra; and
		(c) $L_3$-edge XANES spectrum together with reference spectra of $\mathrm{UF}_{4}$ ($\mathrm{U}^{4+}$) and $\mathrm{UCd}_{11}$ ($\mathrm{U}^{3+}$).
		Panel (a) is reprinted from \cite{UTe2_O_RIXS} with permission, Copyright (2022) by the American Physical Society; 
		panel (b) is reproduced from \cite{UTe2_XMCD} under \href{https://creativecommons.org/licenses/by/4.0/}{CC BY 4.0} and with permission from Springer Nature, and panel (c) is reproduced from \cite{UTe2_L3_XANES} under the same license and with permission from AAAS.
		}
		\label{fig:UTe2_XANES}
\end{figure*}
Figure~\ref{fig:UTe2_XANES}(a) shows the O-edge XAS spectra of \UTeTe measured at various temperatures, together with atomic multiplet calculations for the $5f^2$ and $5f^3$ configurations.
The spectra were obtained using two photon polarizations, horizontal (HP) and vertical (VP).
The $O$-edge XAS exhibits a characteristic splitting into two main regions ($97$--$102~\mathrm{eV}$ and $104$--$118~\mathrm{eV}$), predominantly arising from the $5d$--$5f$ exchange interaction.
The experimental spectra are compared with atomic multiplet calculations for both $5f^2$ and $5f^3$ initial-state configurations, as shown in the right panel of Fig.~\ref{fig:UTe2_XANES}.
The $5f^3$ calculation produces markedly different energy splittings and branching ratios from those observed experimentally, indicating that the ground state of \UTeTe is better described by a $5f^2$-based configuration.
This interpretation is further discussed in the analysis of $O$-edge RIXS presented in a later subsection.

\subsubsection{\texorpdfstring{\Uedge{M_{4,5}}-edge XANES and XMCD}{U M4,5-edge XANES and XMCD}}
Figure~\ref{fig:UTe2_XANES}(b) shows the \Uedge{M_{4,5}}-edge XANES spectrum of \UTeTe~\cite{UTe2_XMCD}.
The \Uedge{M_{4,5}} XANES spectra of \UTeTe are characterized by broad and relatively featureless line shapes, indicating only weak sensitivity to fine details of the electronic structure at the \Uedge{M_{4,5}} edges.
Nevertheless, the $5f$ electron occupancy can be estimated from the branching ratio
$B = I_{M_5}/(I_{M_4}+I_{M_5})$,
where $I_{M_5}$ and $I_{M_4}$ denote the integrated intensities of the $M_5$ and $M_4$ edges, respectively, by applying the spin–orbit sum rule~\cite{SOsumrule}.
For \UTeTe, the branching ratio is determined to be $B = 0.706$, which lies between the intermediate-coupling values of 0.729 and 0.686 calculated for the pure $5f^3$ and $5f^2$ configurations, respectively~\cite{SOsumrule2}.
By further assuming a reasonable value for the spin–orbit interaction per $5f$ hole in the ground state, this analysis yields an estimated $5f$ occupation of $n_{5f} = 2.57$, consistent with an intermediate-valence character of the \Uf electrons.

In addition, XMCD measurements at the $M$ edge have been performed, and comparisons with macroscopic magnetic measurements suggest a slightly larger $5f$ count of approximately $n_{5f} \simeq 2.8$.
Taken together, the analyses of the $M_{4,5}$-edge XANES and XMCD spectra indicate that the $5f$ occupation in \UTeTe lies in the range $n_{5f} \simeq 2.6$–$2.8$.
It should be noted that quantitative determinations of absolute $5f$ occupation numbers from branching ratios and XMCD rely on model assumptions and parameter choices, which can introduce systematic uncertainties~\cite{SR_Actinides}.
Within these considerations, the $M$-edge spectroscopic results provide strong and consistent support for a pronounced mixed-valence character of the \Uf electrons in \UTeTe.

\subsubsection{\texorpdfstring{\Uedge{L_3}-edge XANES}{U L3-edge XANES}}
Figure~\ref{fig:UTe2_XANES}(c) shows the \Uedge{L_3}-edge XANES spectrum of \UTeTe in comparison with those of a $\mathrm{U}^{4+}$ reference compound ($\mathrm{UF}_{4}$) and a $\mathrm{U}^{3+}$ reference compound ($\mathrm{UCd}_{11}$)~\cite{UTe2_L3_XANES}.  
Although a simple chemical-shift interpretation of the \Uedge{L_3} edge is not strictly applicable to metallic systems such as \UTeTe, the white-line peak position is found to lie between those of the two reference compounds, which is naturally interpreted as reflecting a mixed-valence character of the \Uf electrons.
The spectrum of \UTeTe is also considerably broader than that of the ionic compounds, making a direct valence assignment nontrivial as pointed out later.
Notably, the \Uedge{L_3} line shape closely resembles that of intermetallic uranium systems such as $\mathrm{UCu}_5$, $\mathrm{UNi}_5$, and $\alpha-\mathrm{U}$~\cite{U_XANES}, suggesting an itinerant and mixed-valence character of \Uf states in \UTeTe.

\subsection{RIXS}
Resonant inelastic X-ray scattering (RIXS) is a second-order photon-in--photon-out spectroscopic technique in which an incident X-ray resonantly excites a core electron into an unoccupied state, followed by radiative decay of the intermediate state that emits a photon with reduced energy.
By analyzing the energy loss of the scattered photon, RIXS provides access to a wide range of low- to high-energy excitations, including crystal-field and spin–orbit excitations, collective modes, and charge-transfer processes, many of which are not directly accessible by conventional XAS~\cite{RIXS_deGroot}.
RIXS measurements can be broadly classified into two categories: valence-to-core RIXS and core-to-core RIXS, distinguished by the nature of the decay process and the resulting final state.

In valence-to-core RIXS, the decay involves a valence electron refilling the core hole created in the excitation step, such that the final state closely resembles the ground-state valence configuration.
As a result, this type of RIXS is particularly sensitive to the multiplet structure associated with the dominant electronic configurations in the ground state.

In contrast, in core-to-core RIXS, the decay occurs between two core levels, leaving a core hole in the final state.
Because the final state is strongly influenced by the presence of this core hole, the resulting spectra do not directly resolve the ground-state multiplet structure and are instead often interpreted in terms of white-line positions and overall spectral shapes.
Consequently, core-to-core RIXS requires explicit consideration of core-hole effects, in contrast to valence-to-core RIXS, which more directly reflects intrinsic valence electronic properties.

For uranium compounds, RIXS has been performed at the $O$, $N$, $M$, and $L$ absorption edges, and the current status of the technique and its applications to actinide materials are comprehensively reviewed in Refs.~\cite{SR_Actinides,AC_M_RIXS,UGa2_RIXS,AC_RIXS,M45_RIXS}.  
To date, RIXS studies have been concentrated primarily on uranium oxides, with only a limited number of investigations reported for intermetallic compounds.  
In the case of \UTeTe, RIXS measurements have been carried out at the $O$ ($5d$) edge~\cite{UTe2_O_RIXS} and at the $M_{4,5}$ ($3d$) edges~\cite{UTe2_M_RIXS}.

Owing to its element selectivity, sensitivity to multiplet interactions, and capability to probe a broad spectrum of electronic excitations, RIXS has emerged as a powerful and indispensable tool for investigating correlated $5f$ states in uranium-based materials, including \UTeTe.

\subsubsection{\Uedge{O}-edge RIXS}
\begin{figure*}
	\centering
	\includegraphics[scale=0.45]{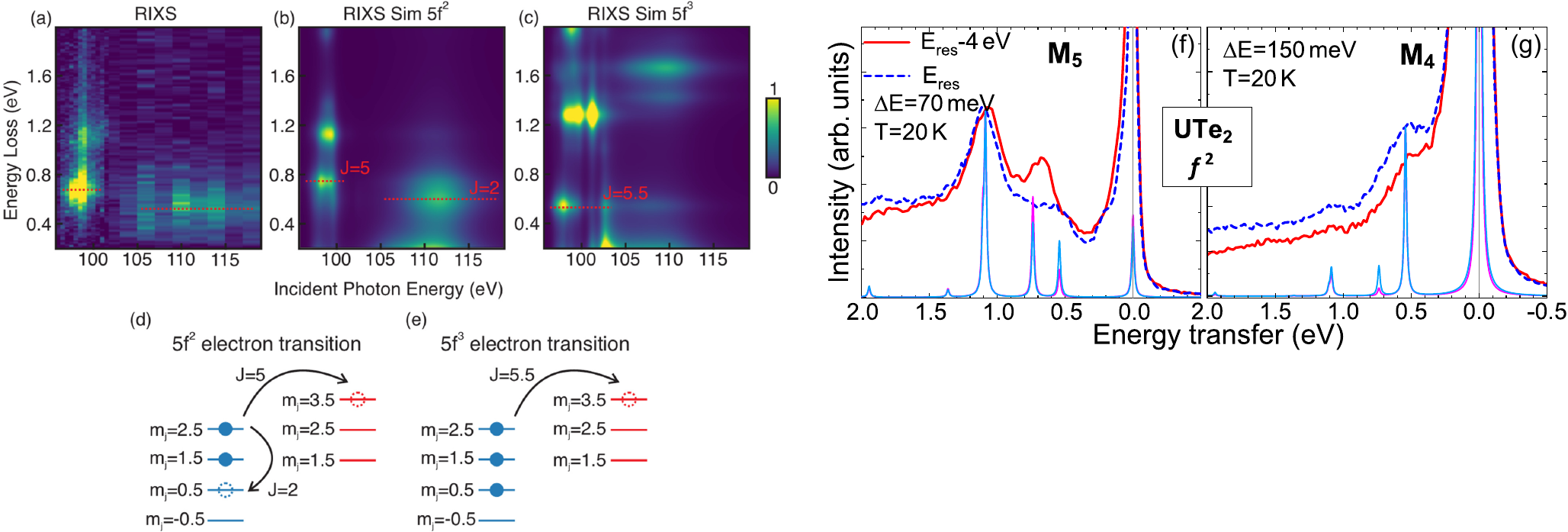}
	\caption{RIXS spectra of \UTeTe.
	(a) \Uedge{O}-edge RIXS spectra of \UTeTe.
	(b), (c) Multiplet simulations for the $5f^2$ and $5f^3$ configurations, respectively.
	(d), (e) Electronic transitions associated with the $5f^2$ and $5f^3$ ground states.
	(f), (g) \Uedge{M}-edge RIXS spectra of \UTeTe.
	Panels (a-e) are reproduced from \cite{UTe2_O_RIXS} with permission, Copyright (2022) by the American Physical Society;
	panels (f,g) are reprinted from \cite{UTe2_M_RIXS} under \href{https://creativecommons.org/licenses/by/4.0/}{CC BY 4.0} with permission, Copyright (2024) by the American Physical Society.
	}
	\label{fig:UTe2_RIXS}
\end{figure*}
Figure~\ref{fig:UTe2_RIXS} compares the experimental \Uedge{O}-edge RIXS spectra of \UTeTe with atomic multiplet calculations for the $5f^{2}$ and $5f^{3}$ configurations~\cite{UTe2_O_RIXS}.
As shown in Fig.~\ref{fig:UTe2_RIXS}(a), the spectra exhibit resonant excitations at approximately $0.7~\mathrm{eV}$ near \hn{\sim99} and at approximately $0.5~\mathrm{eV}$ near \hn{\sim110}.
These features are quantitatively reproduced by the $5f^{2}$ multiplet calculation and can be assigned to intra-atomic $5f^{2}$ excitations with total angular momentum $J=2$ and $J=5$, as summarized in Fig.~\ref{fig:UTe2_RIXS}(d).
In contrast, the multiplet simulation for the $5f^{3}$ configuration shown in Fig.~\ref{fig:UTe2_RIXS}(c) fails to capture the observed excitation energies and spectral weight.
Based on a comparison with atomic multiplet calculations, the \Uedge{O}-edge RIXS spectra are suggestive of a predominantly localized $5f^{2}$ configuration.
The interpretation of this result is discussed in Sec.~\ref{sec:discussion_X-ray}.

\subsubsection{\Uedge{M}-edge RIXS}	
Figure~\ref{fig:UTe2_RIXS} (f) and (g) show the \Uedge{M_{4,5}}-edge RIXS spectra of \UTeTe\ measured under resonant and non-resonant conditions, together with full atomic multiplet calculations based on a $5f^2$ initial configuration.
Both spectra are dominated by a strong elastic peak at $0~\mathrm{eV}$, followed by inelastic features attributed to $5f^n_{\mathrm{GS}}$-type excitations and a broad shoulder associated with the $5f^{n+1}_j \underline{L}_j$ final state.
These characteristic structures are largely reproduced by the multiplet calculations based on $5f^2$ configuration.
However, the experimental spectra are significantly broader than the calculated ones, indicating that the intrinsic multiplet features are smeared out by interatomic interactions and hybridization effects not captured in the purely atomic model.

Assuming a strict $5f^2$ configuration would imply that uranium is formally in a $\mathrm{U}^{4+}$ state, which in turn requires each Te ion in \UTeTe\ to be formally $\mathrm{Te}^{2-}$.
Such a $\mathrm{Te}^{2-}$ anion has a relatively large ionic radius of $2.21~\mbox{\AA}$, incompatible with the short Te(2)--Te(2) chain distance of $3.0355~\mbox{\AA}$ along the $b$ axis.
To reconcile these problems, one assumes that the \orb{U}{6d} orbital accommodates the additional electron released from Te(2).
Consequently, uranium is better described by a $\mathrm{U}~5f^2 6d^1$ configuration rather than a purely $5f^2$ state.
This interpretation is also consistent with the results of the \Uedge{O}-edge RIXS analysis.

\subsection{RXES}
\begin{figure*}
	\centering
	\includegraphics[scale=0.45]{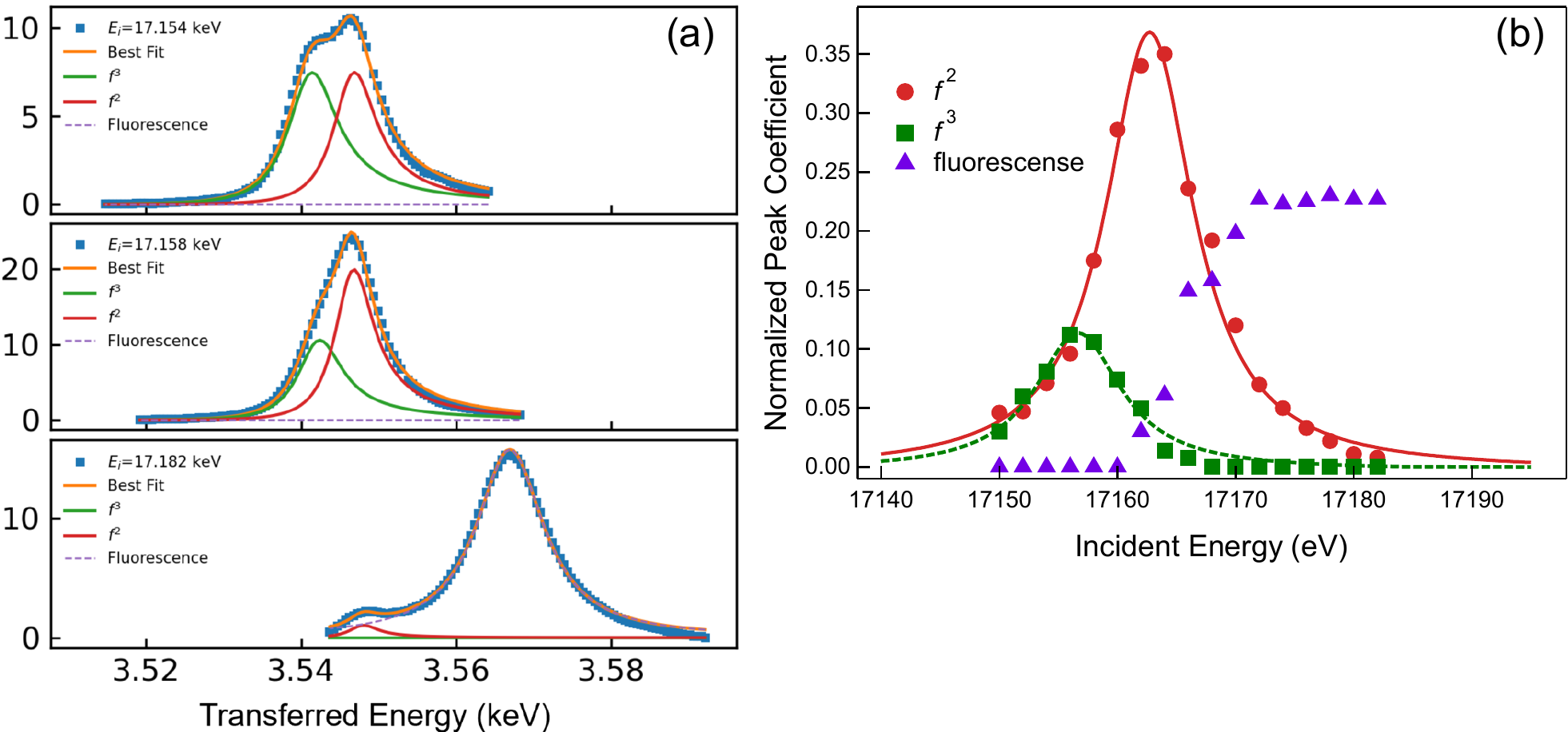}
	\caption{\Uedge{L_3}-edge RXES spectrum of \UTeTe at 2.4~GPa.
	(a) multiconfiguration peak deconvolution for $L_{\alpha1}$ RXES at various incident photon energy $E_i$.
	(b) Normalized peak coefficient data potted vs $E_i$.
	Reprinted from \cite{UTe2_RXES} with permission, Copyright (2024) by the American Physical Society.
	}
	\label{fig:UTe2_L3_XES}
\end{figure*}
Resonant X-ray emission spectroscopy (RXES) is a photon-in--photon-out technique closely related to RIXS, in which the incident photon energy is tuned across an absorption edge while monitoring the emitted fluorescence.
RXES effectively resolves the two-step excitation–decay process by recording the emitted photon energy as a function of the incident energy, providing a two-dimensional intensity map that separates intermediate-state and final-state contributions.
Compared with conventional XANES or fluorescence-yield XAS, RXES offers improved spectral resolution and reduced lifetime broadening, making it particularly powerful for probing mixed valence, core-hole screening, and hybridization effects in correlated-electron systems~\cite{RXES_review}.

At the \Uedge{L} edges, especially the $L_3$ edge (\hnk{\sim 17}), RXES proceeds through a $2p_{3/2} \rightarrow 6d$ excitation, followed by radiative decay typically involving $3d \rightarrow 2p$ or $4d \rightarrow 2p$ transitions.
Because the intermediate state contains a $2p$ core hole screened by both $6d$ and partially occupied $5f$ electrons, the incident-energy ($E_i$) dependence of the RXES map carries detailed information about $5f$–$6d$ hybridization and the effective valence of uranium.
In particular, the position and width of the resonantly enhanced emission features can be used to distinguish different $5f^n$ configurations \cite{AC_RXES}, even in cases where the $L_3$-edge XANES spectrum alone is difficult to interpret due to the metallic or intermetallic nature of the compound.

For \UTeTe, \Uedge{L_3}-edge resonant X-ray emission spectroscopy (RXES) has been employed to refine estimates of the uranium $5f$ occupation number under high pressure~\cite{UTe2_RXES}.
Figure~\ref{fig:UTe2_L3_XES}(a) shows the \Uedge{L_3}-edge RXES spectra measured at several incident photon energies $E_i$ under a pressure of 2.4~GPa.  
The spectra are decomposed into three Lorentzian components, assigned to the $5f^{3}$-derived feature, the $5f^{2}$-derived feature, and a normal fluorescence contribution.  
The corresponding multiconfigurational orbital weights extracted from this analysis are shown in Fig.~\ref{fig:UTe2_L3_XES}(b) as a function of $E_i$.
From these weights, the $5f$ occupation number can be estimated as $n_{5f} = (2I(5f^{2}) + 3I(5f^{3}))/(I(5f^{2}) + I(5f^{3}))$, where $I(5f^{2})$ and $I(5f^{3})$ denote the integrated intensities of the $5f^{2}$ and $5f^{3}$ components, respectively.
Using this procedure, the $5f$ occupation at the lowest pressure accessible in the experiment, 1.8~GPa, is estimated to be $n_{5f} \simeq 2.26$.  
This value provides clear evidence for a mixed-valence ground state, while indicating that the electronic configuration remains relatively close to the $5f^{2}$ limit.  
Although RXES data at ambient pressure were not reported in that study, the pronounced increase of $n_f$ observed upon decreasing pressure in their data suggests that the ambient-pressure $5f$ occupation is likely somewhat larger than this value.  
It should further be noted that $L_3$-edge RXES spectra are strongly influenced by $5f$–$6d$ hybridization effects~\cite{SR_Actinides}, which complicate a direct and quantitative determination of the $5f$ valence.  
Accordingly, the most robust implication of the $L_3$-edge RXES results is the presence of a strong admixture of $5f^{2}$ and $5f^{3}$ configurations in the ground state, rather than a strictly localized valence.

\subsection{\texorpdfstring{\Uff core-level XPS}{U 4f core-level XPS}}
\begin{figure*}
	\centering
	\includegraphics[scale=0.45]{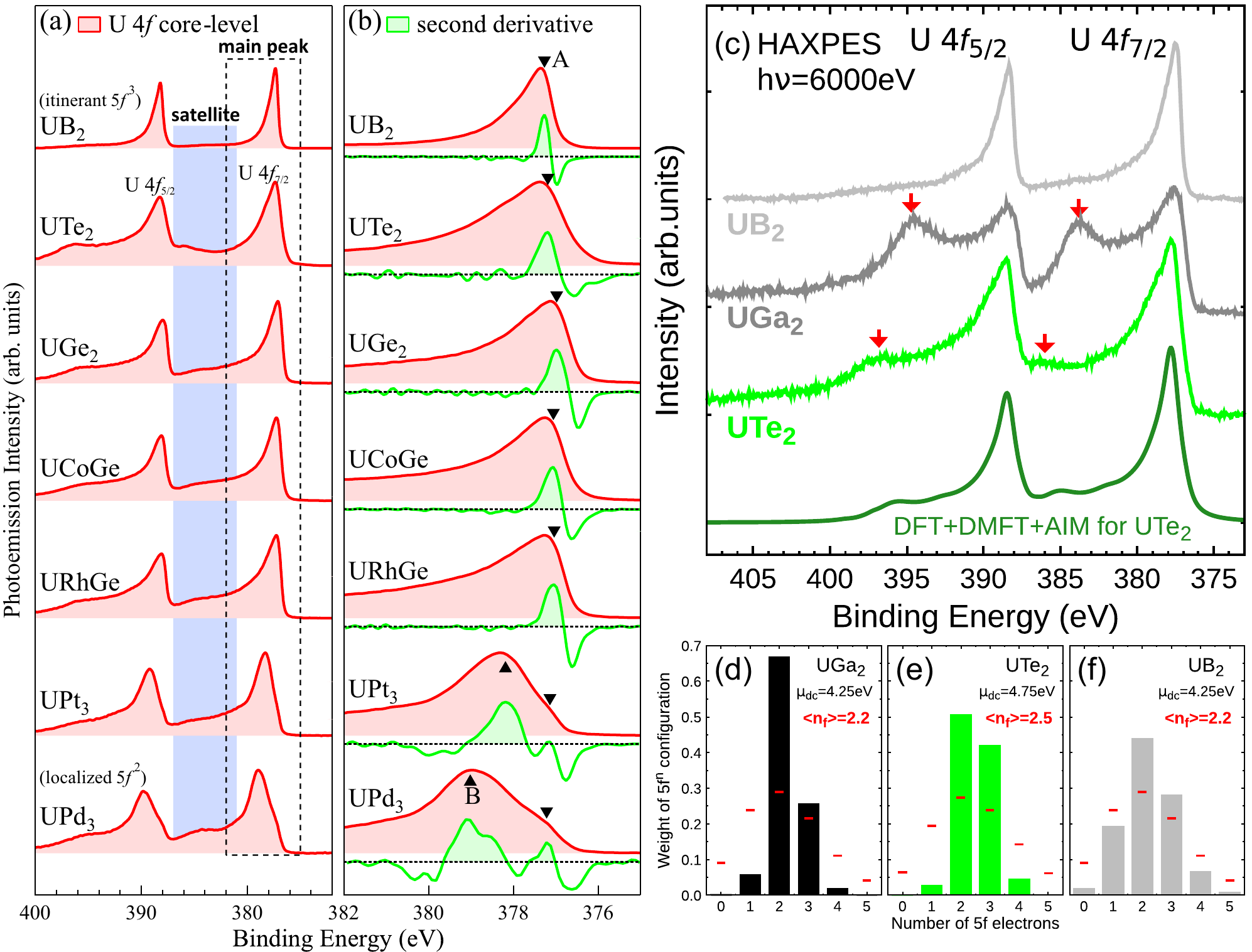}
	\caption{
		\texorpdfstring{$\mathrm{U}~4f$}{U 4f} core-level spectra of \UTeTe and reference uranium compounds.
		(a) \texorpdfstring{$\mathrm{U}~4f$}{U 4f} core-level spectra of \UB, \UTeTe, \UGe, \UCoGe, \URhGe, \UPt, and \UPd measured at \hn{=800}.
		(b) Expanded view of the \texorpdfstring{$\mathrm{U}~4f_{7/2}$}{U 4f7/2} main peaks and their corresponding negative second derivatives.
		(c) \Uff core-level spectra of \UTeTe, $\mathrm{UB}_{2}$, and $\mathrm{UGa}_{2}$ measured at \hn{=6000}, together with the calculated spectrum based on a DFT+DMFT Anderson impurity model (AIM), shown at the bottom.
		(d–f) Valence histograms obtained from DFT+DMFT calculations for these compounds; the corresponding binomial distributions are indicated by red ticks.
		Panels (a) and (b) are reproduced from Fig.~1 of Fujimori \etal, 2021~\cite{UTe2_core}.
		Panels (c--f) are reprinted from \cite{UTe2_U4fDMFT} under \href{https://creativecommons.org/licenses/by/4.0/}{CC BY 4.0}, Copyright (2025) by the American Physical Society.
		}
	\label{fig:UTe2_U4fcore}
\end{figure*}

Core-level XPS is a powerful experimental technique to probe the valence states of atoms in solid. 
In particular, in strongly-correlated materials such ad $d$ and $f$-electron system, there appear satellite structures reflecting complex many body effects in $d$- and $f$-orbitals~\cite{Kotani_review}.
In uranium compounds, complex satellite structure has been observed, which reflect valence state and complex many-body effect of $5f$ orbitals \cite{Ucore,SF_review_JPSJ,ACcore_review}.
The \Uff core-level XPS study has been reported by using two different photon energies: \hn{=800} \cite{UTe2_core} and \hnk{=6} \cite{UTe2_U4fDMFT}.

Figure~\ref{fig:UTe2_U4fcore} (a) shows the \Uff core-level spectrum of \UTeTe together with typical uranium compounds (itinerant compound \UB and localized compound \UPd) and related uranium superconductors \UGe, \UCoGe, \URhGe, and \UPt.
Their negative second derivatives of the \orb{U}{4f_{7/2}} spectra are shown in Fig.~\ref{fig:UTe2_U4fcore} (b) to indicate the peak position.
The \Uff core-level spectra of \UTeTe exhibit a dominant main peak at \EB{=377-379} accompanied by a broad satellite at higher binding energy.
Such structures reflect a superposition of final states with distinct local \Uf configurations, as confirmed by measurements on the diluted alloy compound $\mathrm{U_{0.1}La_{0.9}Pd_2Al_3}$\cite{Fujimori_SSC}.
The main peak energies of \UTeTe, \UGe, \UCoGe, and \URhGe cluster tightly at \EB{=377-377.3}, closely matching itinerant \UB and clearly differing from the more localized \UPd.
This alignment signals that their dominant final-state configurations, and therefore their predominant ground-state \Uf characters, are similar to those of \UB.
The broader lineshapes in \UTeTe and in the ferromagnetic superconductors may contain a weak unresolved \UPd-like contribution at higher binding energy, while the presence of a satellite feature is consistent with partial $5f^2$ character.

Taken together, the data support a mixed-valence ground state in \UTeTe, primarily $5f^3$ with a finite $5f^2$ component.
This picture accords with SX-ARPES, where itinerant QP bands and an incoherent peak have been observed \cite{UTe2_ARPES}.
Moreover, the striking similarity of the \Uff spectra among \UTeTe, \UGe, \UCoGe, and \URhGe reinforces the view that \UTeTe belongs to the family of uranium compounds with itinerant yet strongly correlated \Uf electrons \cite{URhGe_ARPES,UGe2_UCoGe_ARPES,SF_review_JPSJ}.

There are several theoretical approaches to interpret the core-level spectral shape of uranium compounds \cite{UTe2_U4fDMFT,Okada_core,Okada_core_AIM,Zwicknagl_core}.
Among these, a notable advance is the approach based on the Anderson impurity model combined with DFT+DMFT (DFT+DMFT+AIM)~\cite{UTe2_U4fDMFT}.
Figure~\ref{fig:UTe2_U4fcore} (c) shows the \Uff core-level spectra of \UB, \UGa, and \UTeTe measured at \hn{=6000}, and the result of DFT+DMFT+AIM calculation for \UTeTe.
It should be noted that the spectrum measured at \hn{=6000} is almost identical to that measured at \hn{=800} shown in Fig.~\ref{fig:UTe2_U4fcore} (a).
The parameters $U$ and $J$ are listed in table~\ref{calc}.
The characteristic feature of the spectra, asymmetric main peak and weak satellite located at about 8~eV higher than the main peak, are well reproduced by the calculation.
Figure~\ref{fig:UTe2_U4fcore} (d-f) show the valence histograms  from DFT+DMFT of \UGa, \UTeTe, and \UB, respectively, and binomial distribution, which correspond to the fully itinerant, noninteracting bandlike limit for the average filling given by DFT+DMFT.
In \UTeTe, the $5f^2$ configuration is most prominent, while the $5f^3$ configuration is only slightly less pronounced; this yields an average $5f$-shell occupancy of $n_f=2.5$, suggesting that the \Uf states in \UTeTe are in a strongly mixed-valence state.

\section{\label{sec:discussion}Discussion}
\subsection{\texorpdfstring{ARPES perspective on the $5f$ electronic structure}{ARPES perspective on the 5f electronic structure}\label{sec:discussion_ARPES}}
Here we synthesize key insights from ARPES studies of \UTeTe, focusing on (i) the origin of the markedly different spectral features observed in SX-ARPES and low-$h\nu$ ARPES and (ii) the unresolved question of an additional closed three-dimensional (3D) Fermi-surface pocket.
We first summarize the bulk-sensitive picture from SX-ARPES, then rationalize the surface sensitivity of low-$h\nu$ ARPES in terms of the probing depth, and finally discuss how these considerations bear on topology and on the proposed 3D pocket.

SX-ARPES, performed at photon energies of several hundred eV, is relatively bulk sensitive and reveals well-defined, dispersive quasiparticle bands with substantial \Uf spectral weight, together with a pronounced renormalization and an incoherent peak.
These features indicate that the \Uf electrons participate in itinerant quasiparticle states while remaining strongly correlated.

In contrast, low-$h\nu$ ARPES (\hn{\lesssim 100}) is dominated by rapidly dispersing light bands and shows little or no signature of the heavy quasiparticle bands seen in SX-ARPES (Fig.~\ref{fig:UTe2_ARPES_EF})~\cite{UTe2_ARPES_Wray}.
Given that both measurements probe similar high-symmetry directions, the low-$h\nu$ spectra are most naturally interpreted as containing a substantial surface contribution.

This surface sensitivity follows from the short probing depth at low kinetic energies: for $E_{\mathrm{kin}} \lesssim 100~\mathrm{eV}$, the IMFP is less than 5~\AA~\cite{IMFP}, so surface reconstruction and surface-driven localization of \Uf electrons can be emphasized~\cite{SF_review_JPCM}.
By contrast, SX-ARPES at $E_{\mathrm{kin}} \gtrsim 500~\mathrm{eV}$ yields IMFPs exceeding 10~\AA, which is commonly taken as sufficient for bulk sensitivity.
This is consistent with the essentially identical $\mathrm{U}\,4f$ core-level line shapes measured at \hn{=800} and at \hnk{=6}, corresponding to probing depths of roughly 10 and 40~\AA, respectively~\cite{IMFP}.

These surface--bulk differences become particularly important when discussing possible topological superconductivity and associated Majorana boundary modes.
If superconductivity in \UTeTe is topological, Majorana modes are expected as boundary (surface/edge) Andreev bound states, whose existence and visibility depend on the surface Fermi-surface topology, the local symmetry environment, and surface disorder~\cite{UTe2_GGAU,DFTDMFT_Choi,Jiao_UTe2_Nature2020}.
Low-$h\nu$ ARPES suggests that the near-surface electronic structure can approach a \ThTeTe-like Fermiology with reduced \Uf contribution.
Such a modification of the surface electronic structure could, in principle, alter the relevant topological invariants and thus change the boundary spectrum relative to that expected from the bulk.

With this surface--bulk dichotomy in mind, we return to another unresolved issue: the possible existence of a closed 3D Fermi-surface pocket.
Several DMFT-based calculations predict such a pocket either near \Gm or \pnt{Z} (Sec.~\ref{sec:3DFS_calc})~\cite{DFTDMFT_Choi,DFTDMFT_Halloran,UTe2_LQSGWDMFT1,UTe2_LQSGWDMFT2}.
Halloran \ea predict an electron-like pocket near \pnt{Z}, which qualitatively resembles the localized intensity enhancement reported in low-$h\nu$ ARPES Fermi-surface maps.
However, the dispersion inferred from the low-$h\nu$ ARPES feature (a few hundred meV) is far larger than the $\lesssim$ few tens of meV predicted by DMFT, indicating a substantial quantitative discrepancy.
Choi \ea and Kang \ea predict a hole-like pocket centered at \Gm; however, the relevant band lies only a few meV from \EF, well below the typical energy resolution of present ARPES measurements on \UTeTe.

Taken together, the bands responsible for the DMFT-predicted 3D pockets are so shallow that they lie at, or below, the current experimental resolution limit of ARPES.
Therefore, existing ARPES data neither confirm nor definitively rule out the presence of such pockets.
Resolving this issue will require improved energy resolution, enhanced bulk sensitivity, and systematic photon-energy control to tune the probed $k_z$.

Overall, ARPES studies of \UTeTe support a picture in which \Uf electrons form strongly renormalized itinerant quasiparticle bands in the bulk, while surface-sensitive measurements emphasize more localized features that complicate the interpretation of near-\EF states.
The combined use of SX-ARPES and low-$h\nu$ ARPES is therefore essential for disentangling bulk and surface contributions and for clarifying the dual itinerant--localized character of the \Uf electrons in this material.

\subsection{\texorpdfstring{X-ray spectroscopy perspective on the $5f$ electronic structure}{X-ray spectroscopy perspective on the 5f electronic structure}\label{sec:discussion_X-ray}}
As reviewed in Sec.~\ref{sec:Xray}, a wide range of X-ray spectroscopic techniques has been applied to \UTeTe; nevertheless, the uranium valence state inferred from these measurements has not converged on a unique interpretation.  
As summarized in Table~\ref{xrays_table}, analyses based on direct XAS and RIXS measurements are frequently interpreted in terms of a predominantly localized $5f^{2}$ configuration.  
In contrast, core-level photoemission studies consistently indicate an intermediate-valence ground state characterized by a substantial admixture of $5f^{2}$ and $5f^{3}$ configurations.  
Comparable discrepancies have been reported for other uranium intermetallic compounds, such as \URuSi~\cite{URu2Si2_3DARPES}, indicating that this issue is not specific to \UTeTe but instead reflects more general aspects of $5f$-electron spectroscopy in metallic systems.

The origin of these apparently conflicting conclusions lies in the fundamentally different sensitivities of the underlying spectroscopic processes.  
In particular, XAS involving direct $d \rightarrow f$ excitations is known to exhibit only weak chemical sensitivity to the $5f$ valence in intermetallic compounds~\cite{U_XAS_Tobin}.
By contrast, \Uff core-level photoemission spectra display pronounced variations across uranium materials, directly reflecting changes in the local $5f$ electronic configuration as demonstrated in Appendix A.
This contrasting behavior can be rationalized within a minimal two-level model~\cite{XPS_XAS_process}, discussed in Appendix~B, which highlights the decisive role of final-state Coulomb interactions in shaping the spectral response of different X-ray probes.

From this perspective, core-level photoemission emerges as the most valence-sensitive X-ray spectroscopic probe for metallic uranium systems.
For \UTeTe, the experimentally observed \Uff core-level line shape is quantitatively reproduced by DFT+DMFT calculations combined with an Anderson impurity model, yielding a substantial contributions from both $5f^{2}$ and $5f^{3}$ configurations, resulting in $n_f \sim 2.5$.

This estimate is consistent with the pronounced mixed-valence character inferred from $M_{4,5}$-edge XANES and XMCD measurements and is broadly compatible, within experimental and methodological uncertainties, with the RXES result ($n_f \gtrsim 2.26$).
This is also consistent description of both coherent quasiparticle features and incoherent spectral weight in the SX-ARPES spectra.
When these results are considered together, a coherent picture emerges in which the \Uf states of \UTeTe are best described as strongly mixed valent, with substantial contributions from both $5f^{2}$ and $5f^{3}$ configurations.

In contrast, the origin of the apparent discrepancy between this mixed-valence picture and interpretations based on $O$- and $M$-edge RIXS, which often favor a nearly localized $5f^{2}$ electronic structure, remains an open question.
Notably, $O$-edge and $M$-edge RIXS studies of the hidden-order compound \URuSi have been interpreted as evidence for a localized $5f^{2}$ ground state, largely because the measured spectra closely resemble those of reference compounds such as $\mathrm{UO}_2$ and \UPd, which are commonly associated with localized $5f^{2}$ configurations~\cite{AC_M_RIXS,URS_ORIXS}.
These observations indicate that $O$- and $M$-edge RIXS tend to emphasize atomic-like excitations of the \Uf shell when such features are present.

Accordingly, similarities between intermetallic spectra and those of localized reference compounds should be interpreted as reflecting pronounced atomic-like excitation character, rather than as definitive evidence for a purely localized $5f^{2}$ ground state.
Moreover, the shallow probing depth at photon energies of $\sim 100~\mathrm{eV}$ and the relatively weak core-hole potential at the $O$ edge further limit the sensitivity of $O$-edge RIXS to valence fluctuations and hybridization effects~\cite{SR_Actinides}.
These considerations suggest that conclusions regarding the ground-state $5f$ valence drawn from $O$- and $M$-edge RIXS should be interpreted with appropriate caution.

\section{\label{sec:summary}Summary and Outlook}
In this review, we have synthesized recent experimental and theoretical progress on the electronic structure of the unconventional superconductor \UTeTe in normal state, with particular emphasis on how the uranium \Uf states appear across different probes and energy scales.
By integrating ARPES, X-ray spectroscopies, and modern electronic-structure calculations, we aim to provide a coherent framework for interpreting diverse and sometimes seemingly contradictory experimental observations.

ARPES measurements spanning a wide photon-energy range reveal a pronounced dichotomy between surface-sensitive and bulk-sensitive regimes.
Low-$h\nu$ ARPES preferentially highlights relatively light, rapidly dispersing bands with reduced $5f$ spectral weight near~\EF, whereas bulk-sensitive SX-ARPES reveals strongly renormalized heavy quasiparticle bands close to \EF, together with incoherent spectral weight at higher binding energies.
These observations underscore that quantitative discussions of the intrinsic low-energy electronic structure of \UTeTe require careful consideration of the bulk sensitivity of the probe.

Complementary insights are provided by X-ray spectroscopies probing different uranium absorption edges.
Direct $d \rightarrow f$ XAS and RIXS measurements are often broadly consistent with an ionic-like $5f^{2}$ reference, highlighting that their chemical sensitivity can be modest in uranium intermetallic compounds.
In contrast, uranium core-level photoemission supported by DFT+DMFT-based modeling, together with XANES/XMCD and RXES analyses provides clear evidence for valence fluctuations and strong $f$--ligand hybridization.
Taken together, these results indicate that the \Uf states in \UTeTe are in a strong mixed-valence state and cannot be captured within a purely localized or purely itinerant picture.

Overall, the available spectroscopic evidence supports a strongly correlated, mixed-valence normal state with substantial admixture of $5f^{2}$ and $5f^{3}$ configurations.
This framework naturally accounts for the coexistence of heavy quasiparticles and incoherent spectral features and for the pronounced technique dependence of the inferred electronic structure.

Looking forward, several key challenges remain.
(i) Establishing a quantitatively consistent low-energy band structure will require higher-resolution and systematically controlled bulk-sensitive ARPES (including photon-energy dependence to constrain $k_z$), together with improved characterization of cleavage-dependent and surface-reconstructed electronic states.
(ii) The existence of extremely shallow three-dimensional Fermi-surface pockets predicted by DMFT-based approaches remains unresolved and calls for decisive experiments with enhanced resolution and bulk sensitivity, as well as theory that treats realistic surfaces and correlations on an equal footing.
(iii) Connecting the mixed-valence normal state to superconductivity will benefit from measurements that track the evolution of the \Uf spectral weight and valence histogram under controlled tuning parameters (temperature, pressure, and magnetic field), and from theory that links these changes to pairing interactions.

More broadly, \UTeTe provides a benchmark for clarifying how mixed valence and strong spin--orbit coupling, together with surface--bulk differentiation, shape the low-energy electronic structure of uranium-based correlated materials and constrain scenarios for unconventional (and possibly topological) superconductivity.

\ack{
We thankfully acknowledge D. Aoki, Y. Yanase, J. Ishizuka, H. Yamagami, B. Kang, A. Hariki, L. H. Tjeng, and A. Serving for the helpful discussion and providing materials.
The experiment was performed under Proposal No. 2022B3811, 2023A3811, 2023B3811, 2024A3811, 2024B3811 and 2025A3811 at SPring-8 BL23SU.
The present work was financially supported by JSPS KAKENHI Grant Numbers JP26400374, JP16H01084, JP18K03553, JP20KK0061, and JP22H03874.
}

\section*{Appendix A. Weak sensitivity of direct XAS to uranium valence state}
\addcontentsline{toc}{section}{Appendix A. Weak sensitivity of direct XAS to uranium valence state}
\begin{figure*}
	\centering
	\includegraphics[scale=0.5]{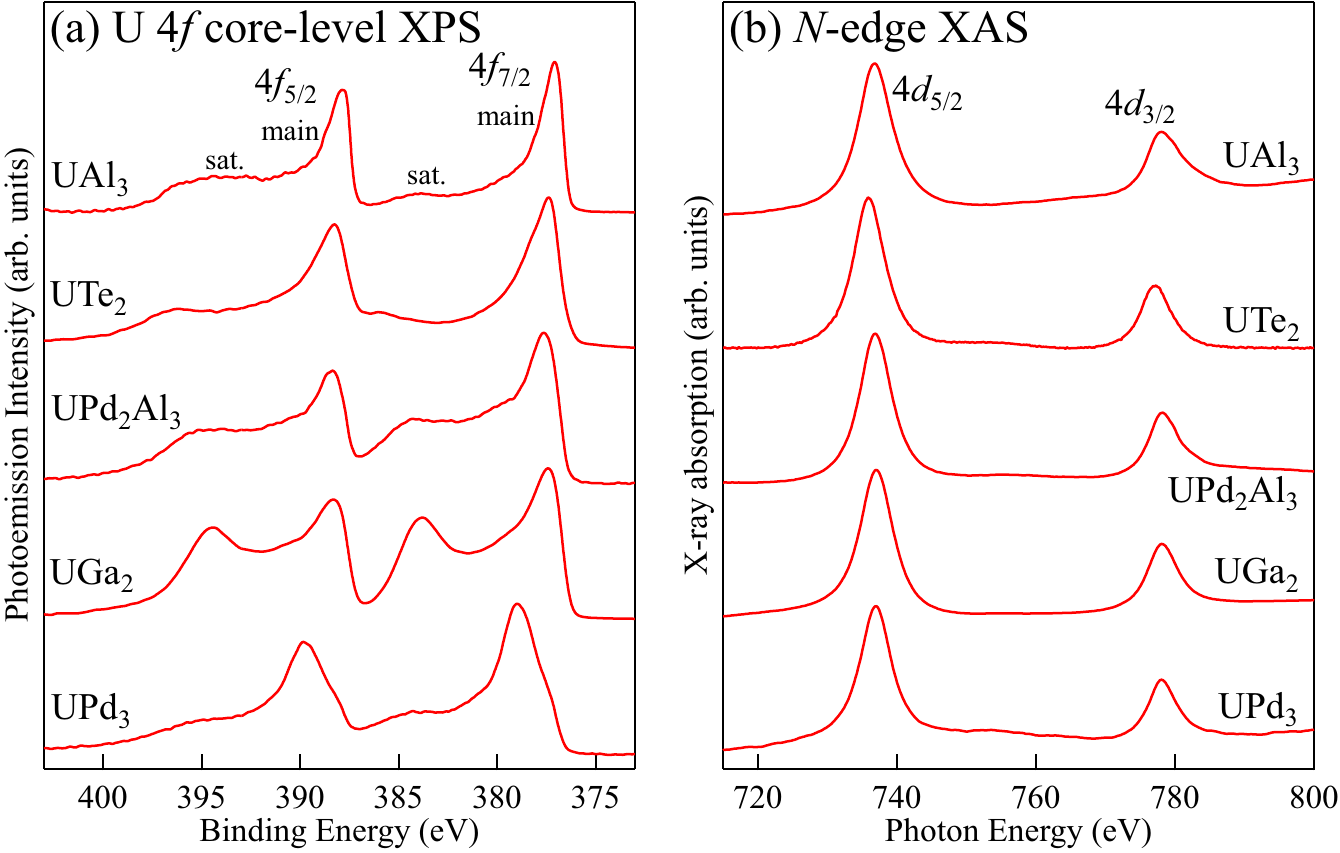}
	\caption{Comparison of (a) \Uff core-level XPS spectra and (b) $N_{4,5}$-edge XAS spectra for the metallic uranium compounds $\mathrm{UAl}_3$, $\mathrm{UTe}_2$, $\mathrm{UPd_2Al_3}$, $\mathrm{UGa}_2$, and $\mathrm{UPd}_3$.
}
	\label{fig:4fXPS_NXAS}
\end{figure*}
\begin{figure*}
	\centering
	\includegraphics[scale=0.45]{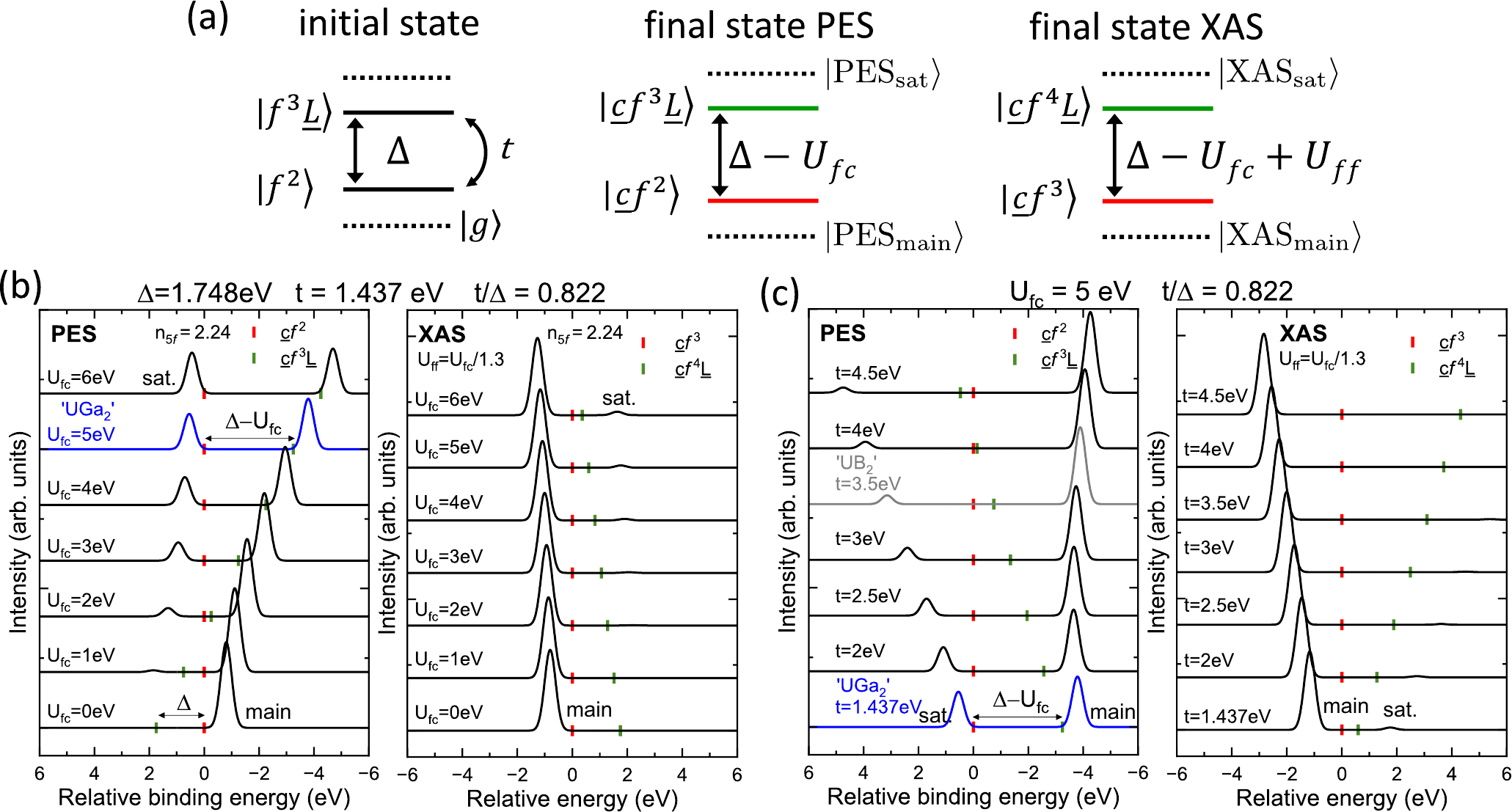}
	\caption{(a) Schematic energy diagram of the initial (ground) state, PES final state, and XAS final state.
	(b) The role of $U_{fc}$ in the core-level photoemission and XAS spectra in the two-level model.
	(c) The role of the hybridization $t$ in the core-level photoemission and XAS spectra.
	Reprinted from \cite{UTe2_U4fDMFT} under \href{https://creativecommons.org/licenses/by/4.0/}{CC BY 4.0} with permission, Copyright (2024) by the American Physical Society.
	}
	\label{PES_XAS_model}
\end{figure*}
To illustrate the limited sensitivity of direct XAS to the uranium valence state, we compare the \Uff core-level XPS spectra with the $N_{4,5}$-edge XAS spectra of \UTeTe\ and representative uranium compounds.  
Figure~\ref{fig:4fXPS_NXAS} presents a comparison of (a) the \Uff core-level XPS spectra and (b) the $N_{4,5}$-edge XAS spectra of \UTeTe with those of the itinerant compound \UAl, the heavy-fermion compound \UPdAl, and the more localized compounds \UGa and \UPd.

The \Uff core-level XPS spectra shown in Fig.~\ref{fig:4fXPS_NXAS} (a) display pronounced compound-dependent changes, both in the line shape of the main peak and in the relative intensity and position of satellite features, directly reflecting differences in the \Uf valence state.
In contrast, the $N_{4,5}$-edge XAS spectra shown in Fig.~\ref{fig:4fXPS_NXAS}(b) are dominated by a single white-line feature and show essentially no variations among these compounds; notably, a very similar spectral shape is also observed in $\mathrm{U}_3\mathrm{O}_8$, an extreme case with a localized $5f^1$ configuration~\cite{U3O8_N_XAS}.
This insensitivity demonstrates that direct $d \rightarrow f$ XAS provides only weak discrimination of the uranium valence configuration.
In these spectra, the line shape is largely governed by the strong core-hole potential and the resulting atomic-like multiplet structure, which tends to produce a dominant white-line feature that is only weakly perturbed by modest changes in the ground-state valence and hybridization, as demonstrated in Appendix B.
Accordingly, changes in $n_{5f}$ that are readily visible in core-level photoemission as spectral-weight transfer can be largely washed out in direct XAS.
A similarly weak chemical sensitivity has been reported for the uranium $O$- and $M$-edge XAS spectra~\cite{U_XANES}, which remain largely unchanged not only among intermetallic compounds but even between metallic systems and ionic compounds such as oxides and fluorides.
These observations underscore that core-level photoemission provides substantially richer and more direct information on the uranium valence state than direct $d \rightarrow f$ XAS.

\section*{Appendix B. A minimal two-level model for XAS and core-level XPS}
\addcontentsline{toc}{section}{Appendix B. A minimal two-level model for XAS and core-level XPS}
The weak chemical sensitivity of XAS spectra can be understood within a minimal two-level model that captures the essential valence fluctuations of uranium, consisting of $\ket{f^{2}}$ and $\ket{f^{3}\underline{L}}$ configurations in the ground state, as schematically illustrated in Fig.~\ref{PES_XAS_model}(a) \cite{XPS_XAS_process}.  
Within this framework, the energy separation between the two configurations is given by
$\Delta = \epsilon_f - \epsilon_{\mathrm{L}} + 2U_{ff}$,
where $\epsilon_f$ and $\epsilon_{\mathrm{L}}$ denote the bare $5f$-level and ligand-state energies, respectively, and $U_{ff}$ is the on-site Coulomb interaction between $5f$ electrons.  
Hybridization between these configurations is described by the matrix element $t$, which mixes the ionic and charge-transfer states in the ground state.

In direct XAS, the initial configurations $\ket{f^{2}}$ and $\ket{f^{3}\underline{L}}$ are promoted to the corresponding final states $\ket{\underline{c}f^{3}}$ and $\ket{\underline{c}f^{4}\underline{L}}$, respectively, where $\underline{c}$ denotes the presence of a core hole created at the $O$ or $M$ edge.  
The resulting energy separation between these final states is
$\Delta_{\mathrm{XAS}} = \Delta + U_{ff} - U_{fc}$,
with $U_{fc}$ representing the Coulomb interaction between the $5f$ electrons and the core hole.  
Because $U_{ff}$ and $U_{fc}$ have comparable magnitudes in uranium compounds, one typically finds $\Delta_{\mathrm{XAS}} \approx \Delta$.
In this regime, the dipole matrix element for the dominant white-line transition,
$\bra{\mathrm{XAS_{main}}}\hat{f}^{\dagger}\hat{c}\ket{g}$,
is close to unity, whereas that for the satellite transition,
$\bra{\mathrm{XAS_{sat}}}\hat{f}^{\dagger}\hat{c}\ket{g}$,
is strongly suppressed.  
Here, $\ket{g}$ denotes the ground-state wave function, while $\hat{f}$ and $\hat{c}$ are the annihilation operators for $5f$ and core electrons, respectively.  
As a consequence, the XAS final state is dominated by an ionic-like configuration, and satellite features carry negligible spectral weight.  
Thus, for uranium intermetallic compounds, direct $d \rightarrow f$ XAS is often well described by a simple ionic picture based on a formal $5f^{2}$ configuration, with only limited sensitivity to the actual $5f$ valence and the strength of $f$–ligand hybridization.

By contrast, in core-level X-ray photoemission spectroscopy (XPS), the relevant energy separation between the final states $\ket{\underline{c}f^{2}}$ and $\ket{\underline{c}f^{3}\underline{L}}$ is given by
$\Delta_{\mathrm{XPS}} = \Delta - U_{fc}$,
which can differ substantially from $\Delta$.  
This leads to a pronounced redistribution of spectral weight between the main line and satellite structures, rendering core-level photoemission far more sensitive to the underlying $5f$ electronic configuration.
 
These effects are illustrated in Figs.~\ref{PES_XAS_model}(b) and (c).  
Figure~\ref{PES_XAS_model}(b) highlights the role of $U_{fc}$ in the shape of the core-level photoemission and XAS spectra.  
As $U_{fc}$ increases, $\Delta_{\mathrm{XPS}}$ becomes negative, and the case of $U_{fc}=5~\mathrm{eV}$ corresponds to a localized compound such as $\mathrm{UGa}_2$.  
While the photoemission spectra develop pronounced satellite features under these conditions, the corresponding XAS spectra remain largely insensitive to variations in $U_{fc}$.  
Figure~\ref{PES_XAS_model}(c) illustrates the dependence on the hybridization strength $t$.  
With increasing $t$ at fixed $t/\Delta=0.822$, the photoemission spectra exhibit systematic changes in both the position and intensity of the satellite features.  
In contrast, the XAS spectra continue to be dominated by a single broad structure, further emphasizing their reduced sensitivity to hybridization effects. 

\section*{References}
\bibliographystyle{iopart-num}
\bibliography{UTe2_review}

\end{document}